\documentstyle[12pt]{article}
\input psfig
\pssilent
\input html.sty
\begin{document}
\title{MATTERS OF GRAVITY, The newsletter of the APS Topical Group on 
Gravitation}
\begin{center}
{ \Large {\bf MATTERS OF GRAVITY}}\\
\bigskip
\hrule
\medskip
{The newsletter of the Topical Group on Gravitation of the American Physical 
Society}\\
\medskip
{\bf Number 15 \hfill Spring 2000}
\end{center}
\begin{flushleft}

\tableofcontents
\vfill
\section*{\noindent  Editor\hfill}

\medskip
Jorge Pullin\\
\smallskip
Center for Gravitational Physics and Geometry\\
The Pennsylvania State University\\
University Park, PA 16802-6300\\
Fax: (814)863-9608\\
Phone (814)863-9597\\
Internet: 
\htmladdnormallink{\protect {\tt{pullin@phys.psu.edu}}}
{mailto:pullin@phys.psu.edu}\\
WWW: \htmladdnormallink{\protect {\tt{http://www.phys.psu.edu/\~{}pullin}}}
{http://www.phys.psu.edu/\~{}pullin}\\
\hfill ISSN: 1527-3431
\begin{rawhtml}
<P>
<BR><HR><P>
\end{rawhtml}
\end{flushleft}
\pagebreak
\section*{Editorial}

Not much to report here. This newsletter is juicy on new research reports,
which signals good times for our field. Enjoy!
\par
The next newsletter is due September 1st.  If everything goes well this
newsletter should be available in the gr-qc Los Alamos archives under
number gr-qc/0002027. To retrieve it send email to 
\htmladdnormallink{gr-qc@xxx.lanl.gov}{mailto:gr-qc@xxx.lanl.gov}
(or 
\htmladdnormallink{gr-qc@babbage.sissa.it}{mailto:gr-qc@babbage.sissa.it} 
in Europe) with Subject: get 0002027
(numbers 2-8 are also available in gr-qc). All issues are available in the
WWW:\\\htmladdnormallink{\protect {\tt{http://vishnu.nirvana.phys.psu.edu/mog.html}}}
{http://vishnu.nirvana.phys.psu.edu/mog.html}\\ 
A hardcopy of the newsletter is
distributed free of charge to the members of the APS
Topical Group on Gravitation upon request (the default distribution form is
via the web) to the secretary of the Topical Group. 
It is considered a lack of etiquette to
ask me to mail you hard copies of the newsletter unless you have
exhausted all your resources to get your copy otherwise.
\par
If you have comments/questions/complaints about the newsletter email
me. Have fun.
\bigbreak

\hfill Jorge Pullin\vspace{-0.8cm}
\section*{Correspondents}
\begin{itemize}
\item John Friedman and Kip Thorne: Relativistic Astrophysics,
\item Raymond Laflamme: Quantum Cosmology and Related Topics
\item Gary Horowitz: Interface with Mathematical High Energy Physics and
String Theory
\item Richard Isaacson: News from NSF
\item Richard Matzner: Numerical Relativity
\item Abhay Ashtekar and Ted Newman: Mathematical Relativity
\item Bernie Schutz: News From Europe
\item Lee Smolin: Quantum Gravity
\item Cliff Will: Confrontation of Theory with Experiment
\item Peter Bender: Space Experiments
\item Riley Newman: Laboratory Experiments
\item Warren Johnson: Resonant Mass Gravitational Wave Detectors
\item Stan Whitcomb: LIGO Project
\end{itemize}
\vfill
\pagebreak

\section*{\centerline {TGG session in the April meeting}}
\addtocontents{toc}{\protect\medskip}
\addtocontents{toc}{\bf News:}
\addtocontents{toc}{\protect\medskip}
\addcontentsline{toc}{subsubsection}{\it  
TGG session in the April meeting, by Cliff Will}
\begin{center}
    Clifford Will, Washington University, St. Louis\\
\htmladdnormallink{cmw@howdy.wustl.edu}
{mailto:cmw@howdy.wustl.edu}
\end{center}

This year's American Physical Society ``April'' Meeting
will be held in Long Beach, CA, April 29 - May 2.  Your TGG program
committee has organized the following invited and contributed sessions.
The annual TGG business meeting will be held on April 30.  In addition
to our TGG sessions, there will be other interesting sessions,
including one on first results from the Chandra satellite, one on
measurements of G, and 9 general interest plenary talks.  For further
information about the meeting, consult the APS website
\htmladdnormallink{http://www.aps.org/meet/APR00/}
{http://www.aps.org/meet/APR00/}

\bigskip

{\em TGG INVITED SESSIONS}

GRAVITATION IN THE 21ST CENTURY\\
(a report on the recent NRC Decadal Survey
of Gravitational Physics)\\
11:00 a.m., Saturday, April 29
\par
James Hartle
The Future of Gravitational Physics
\par
Peter R. Saulson
Prospects for Gravitational Wave Detection
\par
Saul Teukolsky
Black Hole Physics: Prospects for the Coming Century
\par
Eric G. Adelberger
High-Precision tests of the Gravitational Standard Model
\par
Abhay Ashtekar
Challenges and Opportunities in Quantum Gravity

\bigskip

GAMMA RAY BURSTS: THE CENTRAL ENGINE\\
 (with Division of Astrophysics)\\
2:30 p.m., Saturday, April 29
\par
Chryssa Kouveliotou
Recent developments in gamma-ray burst research
\par
Stan Woosley
Collapsars, Gamma-Ray Bursts, and Supernovae
\par
Maximilian Ruffert
Merging Neutron Star - Black Hole Binaries
\par
Wai-Mo Suen
Numerical Relativity and Neutron Star Mergers
\par
Sam Finn
Detecting gravitational-waves from gamma-ray burst sources
\bigskip

GRAVITY AT SHORT RANGE \\
(joint with Division of Particles and Fields)\\
8:00 a.m., Tuesday, May 2
\par
Nima Arkani-Hamed
Accessible Extra Dimensions
\par
John C. Price
Experiments on Gravitational Strength Forces below 1 cm
\par
Aharon Kapitulnik
Measurements of Gravity at sub-millimeter scales using cantilever technology
\par
Greg Landsberg
Probing Extra Dimensions in Collider Experiments
\par
Lisa Randall
Theoretical Scenarios

\bigskip

{\em TGG CONTRIBUTED SESSIONS AND BUSINESS MEETING}\\
\par
GRAVITATIONAL RADIATION - EXPERIMENT
11:00 a.m., Sunday, April 30
\par
TGG BUSINESS MEETING
5:30 p.m., Sunday, April 30
\par
GRAVITATION: QUANTUM, SINGULAR AND ALTERNATIVE
11 a.m., Monday, May 1
\par
GRAVITATIONAL RADIATION - THEORY AND NUMERICAL RELATIVITY
2:00 p.m., Monday, May 1

\section*{\centerline{ NRC report}}
\addtocontents{toc}{\protect\smallskip}
\addcontentsline{toc}{subsubsection}{\it  
NRC report, by Beverly Berger}
\begin{center}
Beverly Berger, Oakland University\\
\htmladdnormallink{berger@oakland.edu}
{berger@oakland.edu}
\end{center}

The Committee on Gravitational Physics of the National Research
Council has completed its review of the past 10 years of
gravitational physics research and made its recommendations
for the next 10 years. The committee's report,
``Gravitational Physics: Exploring the Structure of Space and Time''
is available from the National Academy Press website at

\htmladdnormallink{http://www.nap.edu/catalog/9680.html}
{http://www.nap.edu/catalog/9680.html}

Be sure to click on ``READ" for the on-line version.

\vfill
\pagebreak

\section*{\centerline {MG9 Travel Grant for US researchers}}
\addtocontents{toc}{\protect\smallskip}
\addcontentsline{toc}{subsubsection}{\it  
{MG9 Travel Grant for US researchers, by Jim Isenberg}}
\begin{center}
Jim Isenberg, University of Oregon\\
\htmladdnormallink{jim@newton.uoregon.edu}
{jim@newton.uoregon.edu}
\end{center}

At the University of Oregon have applied for an NSF grant to fund travel
to MG IX in Rome, Italy, this July 2-8. While the grant has not been
approved, there is a reasonable chance that it will be. People who wish to
apply for these funds should fill out the application form below, and send
it (before 15 March)
\par
Note the following:
1) A US scientist is one who generally works in the US. Nationality
    is not relevant.
2) The funds can only be used to cover transportation to and from
    Rome.  Housing and registration costs must be met otherwise.
3) US carriers, if available, must be used. (You are advised to
    consider making airline reservations as soon as possible to secure a
    good fare)
4) The selection of the people to receive travel support will be
    made by a panel of experienced US scientists (both theorists
    and experimentalists), based on the information on the returned
    application forms.  Preference will be given to those with NSF
    grants and those with no federal research support.
\par
Please disseminate this information to your colleagues.
\par
Application for International Travel Grant Funds for U.S.
Participants in MG IX. 
\par
Name
\par
Address*
\par
e-mail address*
\par
Phone/Fax number*
\par
Present position and home institution
\par
Previous positions and home institutions for the past three years
\par
Date and place of most advanced degree
\par
Area of research
\par
Are you: Plenary speaker - Workshop chair - Lead Author on Contributed paper
\par
List all current grants in gravitational physics
\par
List the international meetings in gravitational physics attended
during the past three years (note if you have received NSF travel funds
since GR14):
\par
Estimate Travel Cost for MG IX
\par
Please add anything else you wish the selection committee to know. If
you think that your work may be unfamiliar to us, please ask a senior
scientist to send us a short letter of recommendation.
\par
Signature\hspace{3cm}        Date
\par
Return this application by 15 March  to
Gayle Asburry:  Dept of Math, Univ of Oregon, Eugene, OR 97403
or e-mail to asburry@math.uoregon.edu
\par
*State address, phone number, etc, for March through June
\vfill
\newpage

\section*{\centerline {
How many coalescing binaries} \\
\centerline{are there waiting to be detected?}}
\addtocontents{toc}{\protect\medskip}
\addtocontents{toc}{\bf Research Briefs:}
\addtocontents{toc}{\protect\medskip}
\addcontentsline{toc}{subsubsection}{\it  
How many coalescing binaries are there?,
by Vicky Kalogera}
\begin{center}
Vassiliki Kalogera, Harvard-Smithsonian Center for Astrophysics\\
\htmladdnormallink{vkalogera@cfa.harvard.edu}
{mailto:vkalogera@cfa.harvard.edu}
\end{center}

\noindent 
 The inspiral and coalescence of close binaries with two compact objects,
neutron stars (NS) or black holes (BH), are considered to be some of the
most important sources of gravitational waves. Assessment of their
detectability is crucial and depends on two factors: (i) The strength of
the inspiral gravitational radiation signal in the frequency range of
interest, which determines the maximum distance ($D_{\rm max}$) out to
which coalescing binaries could be detected given a certain detection
system. For LIGO II (and I), the most recent estimates of $D_{\rm max}$
are reported in the latest version of the LSC White Paper on Detector
Research and Development [1]: 450\,Mpc (20\,Mpc) for NS--NS binaries,
1000\,Mpc (40\,Mpc) for NS--BH binaries, and 2000\,Mpc (100\,Mpc) for
BH--BH binaries (assuming 10\,M$_\odot$ BH). (ii) The rate of coalescence
events out to these maximum distances. This rate depends on our
expectation of the Galactic coalescence rates and their extragalactic
extrapolation. Using the above $D_{\rm max}$ estimates and the method of
extrapolation to galaxies other than the Milky Way developed by Phinney
(1991) [2] (based on the blue-light luminosities associated with galaxy
star formation history), it can be estimated that the Galactic coalescence
rates required for a LIGO II detection rate of 2--3 events per year are
$\sim 10^{-6}$\,yr$^{-1}$ (NS--NS coalescence) and $\sim
10^{-8}$\,yr$^{-1}$ (BH--BH coalescence).
\par
 On the issue of detectability then the main question concerns estimates
of the Galactic coalescence rates derived based on our current
astrophysical understanding of coalescing binaries. This question has
occupied the astrophysics community for about ten years now. A number of
studies have appeared in the literature with a wide range of results that
often create a confusing picture for the outside reader. In this article I
will try to present an up-to-date review focusing on our best current bet
for a coalescence rate estimate and its most important uncertainties.
\par
Purely theoretical coalescence rates can be predicted using population
synthesis models of the formation of coalescing binaries, given an
evolutionary formation path. The basic idea is that an ensemble of
primordial binaries, formed at a rate in accordance with the Galactic star
formation rate, is followed as it evolves through a long sequence of
evolutionary stages, including multiple phases of mass and
angular-momentum losses, stable or unstable mass transfer, supernovae or
stellar collapse events. The details of these physical processes are not
very well understood at present, so a number of assumptions are necessary
to obtain coalescence rate estimates and exhaustive parameter studies are
essential in assessing the robustness of the results. Recent studies
[3],[4],[5],[6] have mainly focused on the effect of kicks imparted to
compact objects at birth, as well other uncertain factors at various
levels of detail. The results obtained by varying the kick magnitudes
solely lie in the ranges $ < 10^{-7}-5\times 10^{-4}$\,yr$^{-1}$, ~$<
10^{-7}-10^{-4}$\,yr$^{-1}$, and ~$< 10^{-7}-10^{-5}$\,yr$^{-1}$, for
NS--NS, NS--BH, and BH--BH coalescence events, respectively. Other
uncertain factors can {\em further} change the estimates by factors of
10--100. Given such wide ranges of predicted rates, it becomes evident
that population synthesis calculations have a rather limited predictive
power and provide fairly loose constraints on coalescence rates.
\par
 The observed sample of NS--NS binaries with coalescence times shorter
than $10^{10}$\,yr consists of only two systems, PSR~B1913+16 and
PSR~B1534+12, but provides us with an alternative way of estimating the
NS--NS coalescence rate. Phinney (1991) [2] and Narayan et al.\ (1991) [7]
obtained the first empirical estimates based on models for radio-pulsar
selection effects and estimates of the lifetimes of the observed systems.
Both studies obtained an estimate of $10^{-6}$\,yr$^{-1}$ assuming a
NS--NS Galactic scale height of 1\,kpc. Since then, the increase of the
Galactic volume covered by radio pulsar surveys and an upward revision of
the distance estimate to PSR~B1534+12 have lead to a reduction of the
NS--NS coalescence rate. On the other hand, upward corrections have been
applied, which account for beaming effects and the faint end of the pulsar
luminosity function.  Recent estimates [8],[9],[10],[11] lie in the range
$6\times 10^{-7}$\,yr$^{-1}$ to $8\times 10^{-6}$\,yr$^{-1}$. I am
currently involved in a study [12] in which the issues of NS--NS scale
height, pulsar lifetimes, beaming, and small-number sample and
faint-pulsar corrections are examined in detail. Our best estimate for the
Galactic coalescence rate is $1-2\times 10^{-5}$\,yr$^{-1}$. Uncertainties
dominated by the faint-pulsar luminosity correction (which is typically
large and uncertain because of the small-number sample of close NS--NS)
could decrease this estimate to $\sim 10^{-6}$\,yr$^{-1}$ or raise it up
to $\sim 10^{-4}$\,yr$^{-1}$. Although a significant uncertainty in the
estimate persists, it is clear that the empirical estimates of the NS--NS
coalescence rate are more robust than those calculated purely
theoretically.
\par
 Recently, a new candidate NS--NS system (PSR~J1141-6545) was discovered
by the ongoing Parkes Multibeam pulsar survey [13]. Although the nature of
the pulsar companion needs confirmation (it could be a white dwarf) and
the associated selection effects have not been modeled yet, a lower limit
to its contribution to the empirical NS--NS coalescence rate can be
estimated based solely on the pulsar lifetime [12]. Unlike the other two
systems, PSR~J1141-6545 is young with a characteristic age of only
1.45\,Myr and its total lifetime is estimated to 30.5\,Myr. Even if it is
the only such pulsar in the Galaxy, this newly discovered system can
contribute to the coalescence rate by at least $\simeq 3\times
10^{-8}$\,yr$^{-1}$. Taking into account all the corrections, a 10-fold
upward revision of the rate would require that 50 to 200 such pulsars
exist in our Galaxy.
\par
 Information about the detectability of coalescing NS--NS systems can also
be obtained if robust limits to the rate can be derived. So far a safe
upper limit of $\sim 10^{-4}$\,yr$^{-1}$ has been derived based on two
different arguments: (i) the absence (until recently) of any young pulsars
in close NS--NS binaries [14],[10] (this upper limit will be increased by
a multiplication factor equal to the estimated number of pulsars similar
to PSR~J1141-6545 in the Galaxy), and (ii) the maximum ratio of the
formation frequencies of coalescing NS--NS and isolated pulsars similar to
those found in NS--NS systems (freed at the second supernova) and an
empirical estimate of the birth rate of such isolated pulsars [15].
\par
 If we compare the estimated coalescence rates to the requirement for a
LIGO II detection rate of 2--3 events per year, then we can expect a
detection rate in the range of 1--10 (based on the more robust empirical
estimates) or even up to $\sim 100$ per year, based on the derived upper
limits. For NS--BH and BH--BH coalescence, we can only rely on purely
theoretical estimates. Despite the large uncertainties (typically 3--4
orders of magnitude), the ranges for their most part lie {\em above} the
requirements for a couple of events detected per year by LIGO II and imply
detection rates of a few up to even 100-1000 per year. For LIGO I, a
simple volume scaling shows that detection of NS--NS inspiral is rather
unlikely, while BH binaries could be detected provided that the upper ends
of the ranges are closer to reality.
\par
 So far we have dealt with coalescing binaries formed in galactic fields.
Formation of coalescing binaries in globular clusters involves a whole
range of very different processes mostly dominated by stellar interactions
and also differs because of the absence of ongoing star formation over
timescales comparable to the lifetimes of these binaries. The contribution
of clusters to NS--NS coalescence has been found to be negligible [2].
However, a recent study [16] examined the formation of BH--BH binaries
with coalescence times shorter than $10^{10}$\,yr and concluded that their
formation rates are quite high possibly leading to LIGO II detection rates
of $\sim 100$ per year (one event per two years for LIGO I).  Although
these predicted rates may be lower because of necessary cosmological
corrections and loss of systems with very short coalescence timescales,
they are still more than encouraging! 
\par
Overall, it seems fair to say that, despite the uncertainties in the rate
estimates, the prospects for gravitational wave detection from the
inspiral of compact binaries appear to be quite promising, especially for
the upgraded LIGO interferometers.
\par
{\em{References:}}
\par
[1]~~Gustafson, E., Shoemaker, D., Strain, K., and Weiss, R.\ 1999,
{\it LSC White Paper on Detector Research and Development} (LIGO-Project
document, September 11).
\par
[2]~~Phinney, E.S.\ 1991, {\it ApJ}, 380, 17.
\par
[3]~~Lipunov, V.M., Postnov, Prokorov, 1997, {\it MNRAS}, 288,
245.
\par
[4]~~Fryer, C.L., Burrows, A., \& Benz, W.\  1998, {\it ApJ}, 496, 333.
\par
[5]~~Portegies-Zwart, S.Z., and Yungel'son, L.R.\ 1998,
{\it A\&A}, 332, 173.
\par
[6]~~Brown, G.E., and Bethe, H.\ 1998, {\it ApJ}, 506, 780.
\par
[7]~~Narayan, R., Piran, T., \& Shemi, S.\ 1991, {\it ApJ}, 379, 17.
\par
[8]~~van den Heuvel, E.P.J., and Lorimer, D.R.\ 1996, {\it MNRAS}, 283,
37.
\par
[9]~~Stairs, I.H., et al.\ 1998, {\it ApJ}, 505, 352.
\par
[10]~~Arzoumanian, Z., Cordes, J.H., Wasserman, I.\  1998, {\it ApJ},
520, 696.
\par
[11]~~Evans, T., et al.\  2000, to appear in the proceedings of
the XXXIVth Rencontres de Moriond on ``Gravitational Waves and Experimental
Gravity", Les Arcs, France.
\par
[12]~~Kalogera, V., Narayan, R., Spergel, D., \& Taylor, J.\ 2000, to be
submitted to ApJ.
\par
[13]~~Manchester, R.N., et al.\  2000, to appear in {\it Pulsar
Astronomy - 2000 and Beyond}, eds.\ N.\ Wex, M.\ Kramer, \& R.\
Wielebinski.
\par
[14]~~Bailes M.\ 1996, in {\it Compact Stars in Binaries}, IAU Symp.\
No.\ 165, eds.\ J. van Paradijs, E.P.J.\ van den Heuvel, and E.\ Kuulkers
(Dordrecht: Kluwer Academic Publishers), 213
\par
[15]~~Kalogera, V., and Lorimer, D.R.\ 2000, {\it ApJ}, 530, in
press, \htmladdnormallink{{\tt{astro-ph/9907426}}}
{http://xxx.lanl.gov/abs/astro-ph/9907426}.
\par
[16]~~Portegies-Zwart, S.F., and McMillan, S.L.W.\  {\it ApJ
Letters}, 528, L17 (2000).

\vfill
\pagebreak

\section*{\centerline {Black hole critical phenomena:  a brief update}}
\addtocontents{toc}{\protect\smallskip}
\addcontentsline{toc}{subsubsection}{\it
Recent developments in black critical phenomena, by Pat Brady}
\begin{center}
Patrick R Brady, University of Wisconsin-Milwaukee\\
\mbox{\ }
\htmladdnormallink{patrick@gravity.phys.uwm.edu}
{mailto:patrick@gravity.phys.uwm.edu}
\end{center}

The analogy between critical phenomena in statistical physics and
interesting dynamical features observed in numerical simulations of
spherical self-gravitating scalar field collapse was introduced by
Choptuik~[1]. Choptuik numerically evolved one parameter $p$ families of
initial data.  For all families, he found a critical value of the
parameter, $p^*$ say.  No black hole formed in evolutions with $p<p^*$ and
black holes always formed in evolutions with $p>p^*$.  Choptuik also
observed two fundamental properties of solutions with $p\simeq p^*$.
First, they exhibited self-similar echoing, later called discrete
self-similarity, which was universal.  Second, the mass of black holes
formed in marginally super-critical collapse obeyed a scaling law
$M_{\mathrm{BH}} \propto |p - p^*|^\gamma$ with $\gamma \simeq 0.37$
independent of the initial data family.  Choptuik speculated that there was
a universal solution which acted as an intermediate attractor when $p=p^*$.
Since this ground-breaking work, a considerable literature has emerged on
critical phenomena in gravitational collapse.
\par
Different matter models, couplings and symmetries were examined in an
effort to understand the extent of the universality and scaling noted
by Choptuik.  In a tremendous tour-de-force, Abrahams and
Evans~[2] explored the collapse of axisymmetric
gravitational wave configurations; they found tentative evidence for
discrete self-similarity and a scaling law for black hole mass with $\gamma
\simeq 0.37$.  Evans and Coleman~[3] considered the
collapse of radiation fluid spheres; they found that near critical
evolutions exhibit continuous self-similarity (CSS), and that the scaling
exponent for the black hole mass is $\gamma \simeq 0.36$.  Evans and
Coleman went further by constructing the CSS solution which is the
intermediate attractor in perfect fluid collapse.  Koike \textit{et
al.}~[4] examined the spectrum of perturbations around this solution and
demonstrated that the solution is one mode unstable.  The critical exponent
$\gamma$ is related to the growth rate $\beta$ of the unstable mode as
$\gamma= 1/\beta$.  The completion of this program, as suggested by Evans
and Coleman, established the origin of the mass scaling law for black hole
mass.  Maison~[5] used the method to argue that the scaling exponent would
be a function of $k$ for equations of state with $P=k \rho$ where $P$ is
pressure and $\rho$ energy density.
\par
Since the last report in \emph{Matters of Gravity}, this field has
consolidated further.  It is now well understood what constitutes a
critical solution for gravitational collapse, and several important
``analytic'' results indicate our understanding of critical phenomena is
correct.  Following the work of Koike \textit{et al.}, Gundlach applied
perturbative techniques to the massless scalar field critical solution
which he computed directly~[6,7].  His computation of the critical exponent
agreed with the experimentally observed value. This, and the work of Koike
\textit{et al.}, are excellent examples of retrodiction; having the answer,
a direct computation was performed to establish the critical exponent.  But
the techniques used to identify critical solutions and to compute the
associated scaling exponents have been applied by Gundlach, Martin-Garcia,
Maison and others to discover new critical solutions and predict associated
critical exponents.  This work was vindicated by subsequent numerical
computation.  For example, analytic calculations predicted a periodic
wiggle superimposed on the mass scaling law for massless scalar field
collapse~[7,8]; the oscillations were found numerically by Hod and
Piran~[8].  Charged scalar field collapse was predicted to behave as
uncharged massless scalar field near the critical point; the charge obeying
a scaling law similar to the mass but with a critical exponent
$\delta=0.88$. This was also confirmed by Hod and Piran~[9].
\par
Researchers have continued to explore the parameter space of solutions for
a variety of matter models bringing a wealth of new phenomenology to light.
Choptuik, Chmaj and Bizon~[10] studied the collapse of SU(2) Yang-Mills
fields.  They found two distinct types of critical behavior.  In Type II
transitions, the critical solution is discretely self-similar and black
holes of arbitrarily small mass can form.  The appearance of Type I
transitions was a new feature in black hole critical phenomena; the
critical solution is the static Bartnik-McKinnon solution and black hole
formation turns on at finite mass.  Type I phase transitions have also been
found for massive scalar fields~[11] and $SU(2)$ Skyrme models~[12].
Further phenomenology has also been identified in studies of the magnetic
Yang Mills fields.  Choptuik, Hirschmann and Marsa~[13] found transitions
between black holes formed in Type I collapse and black holes formed in
Type II collapse; the critical solution is an unstable colored black hole.
\par
Interestingly, numerical confirmation of Maison's predictions were not
forthcoming until late in 1997~[14].  With this came new understanding of
regular self-similar perfect fluid solutions. Folklore had it that no
regular self-similar solutions existed for $k> 0.899$ in the equations of
state $P=k\rho$.  Neilsen and Choptuik found evidence for such solutions in
their collapse simulations, and re-investigated the exact self-similar
solutions.  They found that the nature of the sonic horizon changes when
$k> 0.899$ \emph{but} regular self-similar solutions do exist.  The
surprise of this was emphasized when Neilsen and Choptuik evolved stiff
fluids with $k=1$ and found a CSS critical solution.  Since an irrotational
perfect fluid with $k=1$ can be recast as a scalar field with timelike
gradient, Neilsen and Choptuik had found a scalar field solution with a CSS
critical solution in contrast top Choptuik's original work.  This issue is
under active investigation.
\par
Lack of space prohibits detailed discussion of the many other lines of
research that have been pursued.  Astrophysical implications of formation
of tiny black holes in the early universe have been considered.  Attempts
have been made to understand the semi-classical corrections to Type II
critical phenomena.  The symmetries of the critical solutions led
to the development symmetry seeking coordinates~[15] which might be
useful in other circumstances.  In a mammoth effort, Gundlach,
Martin-Garcia and Garfinkle~[16] have examined small deviations from
spherical symmetry and predicted scaling relations for angular momentum in
Type II transitions.  The interested reader is referred to Gundlach's review
article~[17] for more details.
\par
So what does the future hold.  There is little doubt that axisymmetric (and
ultimately 3-dimensional) collapse simulations will bring a wealth of new
phenomenology.  Significant effort is under way to produce accurate and
robust codes to perform the parameter space surveys that are needed.  Just
as the initial study of massless scalar field collapse required the
introduction of new techniques into numerical relativity, ongoing
research should foster further developments.
\par
{\em{References:}}
\par
{[1]}
M.~W. Choptuik, {\em{ Phys. Rev. Letters}} {\bf 70}, 9--12 (1993).
\par
{[2]}
A.~M. Abrahams and C.~R. Evans, {\em{Phys. Rev. Letters}} {\bf 70}, 2980 (1993).
\par
{[3]}
C.~R. Evans and J.~S. Coleman, {\em{Phys. Rev. Letters}} {\bf
  72}, 1782 (1994).
\par
{[4]}
T.~Koike, T.~Hara, and S.~Adachi,
  {\em{Phys. Rev. Lett.}} {\bf 74}, 5170  (1995),
  \htmladdnormallink{{\tt{gr-qc/9503007}}}
{http://xxx.lanl.gov/abs/gr-qc/9503007}.
\par
{[5]}
D.~Maison, {\em{Phys. Lett. B}} {\bf 366}, 82 (1996).
\par
{[6]}
C.~Gundlach, {\em{Phys.
  Rev. Lett.}} {\bf 75}, 3214  (1995),
  \htmladdnormallink{{\tt{gr-qc/9507054}}}
{http://xxx.lanl.gov/abs/gr-qc/9507054}.
\par
{[7]}
C.~Gundlach, {\em{Phys.
  Rev. D}} {\bf 55}, 695 (1997),
  \htmladdnormallink{{\tt{gr-qc/9604019}}}
{http://xxx.lanl.gov/abs/gr-qc/9604019}.
\par
{[8]}
S.~Hod and T.~Piran,
  {\em{Phys. Rev. D}} {\bf 55}, 440 (1997),
  \htmladdnormallink{{\tt{gr-qc/9606087}}}
{http://xxx.lanl.gov/abs/gr-qc/9606087}.
\par
{[9]}
S.~Hod and T.~Piran, {\em{Phys. Rev. D}} {\bf 55},
  3485  (1997), \\{{\htmladdnormallink\tt
  gr-qc/9606093}}
{http://xxx.lanl.gov/abs/gr-qc/9606093}.
\par
{[10]}
M.~W. Choptuik, T.~Chmaj, P.~Bizon, {\em{Phys. Rev. Lett.}} {\bf 77},
  424 (1996), \htmladdnormallink{{\tt{gr-qc/9603051}}}
{http://xxx.lanl.gov/abs/gr-qc/9603051};
P.~Bizon and T.~Chmaj, {\em{Acta Phys. Polon.}} {\bf B29}, 1071  (1998),
  \htmladdnormallink{{\tt{gr-qc/9802002}}}
{http://xxx.lanl.gov/abs/gr-qc/9802002}.
\par
{[11]}
P.~R. Brady, C.~M. Chambers, and S.~M. C.~V. Goncalves, {\em{Phys. Rev. D}}
{\bf 56}, 6057 (1997), 
\htmladdnormallink{{\tt{gr-qc/9709014}}}
{http://xxx.lanl.gov/abs/gr-qc/9709014}.
\par
{[12]}
P.~Bizon, T.~Chmaj, {\em{Phys. Rev.}}
  {\bf D58}, 041501  (1998), {{\htmladdnormallink\tt
  gr-qc/9801012}}\\
{http://xxx.lanl.gov/abs/gr-qc/9801012}.
\par
{[13]}
M.~W. Choptuik, E.~W. Hirschmann, and R.~L. Marsa, {\em{Phys. Rev.}} {\bf
D60}, 124011  (1999), \htmladdnormallink{{\tt{gr-qc/9903081}}}
{http://xxx.lanl.gov/abs/gr-qc/9903081}.
\par
{[14]}
D.~W. Neilsen and M.~W. Choptuik,  {\em{Class. Quant. Grav.}} {\bf 17},
 761 (2000),
\htmladdnormallink{{\tt{gr-qc/9812053}}}
{http://xxx.lanl.gov/abs/gr-qc/9812053}; P.~Brady
and M.~J. Cai, in {\em{Proceedings of the 8th Marcel Grossman Meeting}}, T.~Piran, ed.
\newblock World Scientific, Singapore, 1999.
\par
{[15]}
D.~Garfinkle and C.~Gundlach, {\em{  Class. Quant. Grav.}} {\bf 16} 4111 (1999),
  \htmladdnormallink{{\tt{gr-qc/9908016}}}
{http://xxx.lanl.gov/abs/gr-qc/9908016}.
\par
{[16]}
C.~Gundlach, ``Critical gravitational collapse of a perfect fluid with
p=k*rho: Nonspherical perturbations'',
\htmladdnormallink{{\tt{gr-qc/9906124}}}
{http://xxx.lanl.gov/abs/gr-qc/9906124};
C.~Gundlach and J.~M. Martin-Garcia,
``Gauge-invariant and coordinate-independent perturbations of
                  stellar  collapse. I: The interior,''
\htmladdnormallink{{\tt{gr-qc/9906068}}}
{http://xxx.lanl.gov/abs/gr-qc/9906068};
D.~Garfinkle, C.~Gundlach, and J.~M. Martin-Garcia, {\em{Phys. Rev.}} {\bf
D59}, 104012 (1999), \htmladdnormallink{{\tt{gr-qc/9811004}}}
{http://xxx.lanl.gov/abs/gr-qc/9811004}.
\par
{[17]}
C.~Gundlach, ``Critical phenomena in gravitational collapse,'' to appear
in Living Reviews in Relativity, 
  \htmladdnormallink{{\tt{gr-qc/0001046}}}
{http://xxx.lanl.gov/abs/gr-qc/0001046}.

\vfill
\pagebreak

\section*{\centerline{Optical black holes?}}
\addtocontents{toc}{\protect\smallskip}
\addcontentsline{toc}{subsubsection}{\it
Optical black holes?, by Matt Visser}
\begin{center}
Matt Visser, Washington University, St. Louis
\mbox{\ }
\htmladdnormallink{visser@tui.wustl.edu}
{mailto:visser@tui.wustl.edu}
\end{center}

The last few years have seen a lot of interest in condensed matter
analogues for classical Einstein gravity. The most well-developed
of these analog models are Unruh's acoustic black holes (dumb black
holes), but attention has recently shifted to the optical realm. The
basic idea is that in a dielectric fluid the refractive index, the
fluid velocity, and the background Minkowski metric can be combined
algebraically to provide an ``effective metric'' that can be used to
describe the propagation of electromagnetic waves.  The most detailed
and up-to-date implementation of this idea is presented in very recent
papers by Leonhardt and Piwnicki [1--4], which are based on a thorough
re-assessment of the very early work of Gordon [5].
\par
To get the flavour of the way the effective metric is set up, start
with a dispersionless homogeneous stationary dielectric with
refractive index $n$ and write the electromagnetic equations of motion
as
\[
\left(-n^2 {d^2\over dt^2} + \nabla^2 \right) F^{ab} =0.
\]
If you write this in terms of the Minkowski metric $\eta^{ab}$ and
dielectric 4-velocity $V^a$, then
\[
\left\{
-n^2 (V^c\;\nabla_c)^2 + [\eta^{cd} + V^c\; V^d ] \; \nabla_c \nabla_d
\right\} F_{ab} = 0.
\]
Now promote the refractive index and 4-velocity to be slowly-varying
functions of space and time. (Slowly varying with respect to the
wavelength and frequency of the EM wave.) The preceding formula {\em{suggests}} that it is possible to write
\[
{1\over\sqrt{-g}} \;
\partial_c \left( \sqrt{-g}\; g^{cd} \; \partial_d  F_{ab} \right) = 0,
\]
with the (inverse) metric being proportional to
\[
g^{ab} \propto \eta^{ab} - (n^2-1) \; V^a \; V^b.
\]
A more detailed calculation confirms this suggestion, and also lets
you fix the overall conformal factor (it's unity, at least in Gordon's
implementation). If you are only interested in ray optics then fixing
the conformal factor is not important. Once you have this effective
metric in hand, applying it is straightforward (even if the physical
situation is unusual).
\par
There are a few tricks and traps:
\begin{enumerate}
\item
The 4-velocity is normalized using the Minkowski metric $\eta_{ab}
\; V^a \; V^b = -1$,
and in this particular subfield it seems to have become conventional
to define $V_a = \eta_{ab} V^b$, so that the index on the 4-velocity
is lowered with the Minkowski metric. (But for everything else you
raise and lower indices using the effective metric.) The metric itself
is then
\[
g_{ab} =
\eta_{ab} - (n^{-2}-1)\; V_a \; V_b =
[\eta_{ab} + V_a \; V_b ] -  n^{-2} \; V_a \; V_b.
\]
\item
The analogy with Einstein gravity only extends to the kinematic
aspects of general relativity, not the dynamic. There is no analog for
the Einstein equations of general relativity and trying to impose the
Einstein equations is utterly meaningless.
\item
If however, instead of using the energy conditions plus the Einstein
equations, you place constraints directly on the Ricci curvature
tensor or Einstein curvature tensor then you can still prove versions
of the focusing theorems.
\item
As in general relativity, the Riemann tensor and its contractions are
still useful for characterizing the relative motion of nearby
geodesics.
\item
The Fresnel drag coefficient can be read off directly from the
contravariant components of the metric. Specifically
\[
g_{0i} = (n^{-2}-1)\;\gamma^2 \; \vec v.
\]
For low velocities $\gamma\approx1$ this implies that the medium drags
the light as though the medium had an effective velocity
\[
\vec v_{\mathrm{eff}} = (n^{-2}-1) \; \vec v.
\]
The effective velocity of the medium is just the ``shift vector'' in
the metric.  The fact that the Fresnel drag coefficient drops out
automatically should not surprise you at all since we are extracting
all this from a manifestly Lorentz invariant formalism, and so you
must get the same result as from the more usual approach based on the
relativistic addition of velocities
\[
c_{\mathrm{dragged}} = { v + (c/n) \over 1 + {v (c/n)\over c^2 } }
\approx {c\over n} +  (n^{-2}-1) \; v + O(v^2).
\]
\item
There are optical analogs of the notions of ``trapped surface'',
``apparent horizon'', ``event horizon'', and ``optical black hole''
that are in exact parallel to those developed for the acoustic black
holes [6,7].
\item
If you somehow arrange an ``optical event horizon'' of this type, then
there is near-universal agreement among the quantum field theory
community that you should see Hawking radiation from this ``optical
event horizon'', this radiation being in the form of a near-thermal
bath of photons with a Hawking temperature proportional to the
acceleration of the fluid as it crosses the horizon [6,7] --- this is
a very exciting possibility, because we would love to be able to do
some experimental checks on Hawking radiation.
\item
You will be able to probe aspects of {\em{semiclassical quantum
gravity}} with this technique, but it won't tell you anything about
quantum gravity itself.  Because an effective metric of this type is
not constrained by the Einstein equations it allows you only to probe
{\em{kinematic}} aspects of how quantum fields react to being placed on
a curved background geometry, but does not let you probe any of the
deeper {\em{dynamical\/}} questions of just how quantum matter feeds into
the Einstein equations to generate real spacetime curvature. Even
though it should be kept in mind that these ``effective metric''
techniques are limited in this sense, they are still a tremendous
advance over the current state of affairs.
\item
I should mention that I believe the original implementation of
Leonhardt and Piwnicki fails to generate genuine black holes, but that
this can be straightforwardly corrected [8]. Despite this technical
issue, which I believe causes problems for the particular toy model
they discussed, it is clear that the basic idea is fine --- it is
possible to form ``optical black holes'' by accelerating a dielectric
fluid to superluminal velocities (superluminal in the sense
$c/n$). Any region of superluminal fluid flow will be an ergo-region,
and any surface for which the inward component of the fluid flow is
superluminal will be a trapped surface.
\end{enumerate}

Finally, let me emphasize the fundamental experimental reason this is
now all so interesting: experimental physicists have now managed to
get refractive indices up to $n\approx 30,000,000$ which corresponds to
$c/n \approx 10$ meters/second [9] --- and it is this experimental
fact that holds out the hope for doing laboratory experiments in the
not too distant future.
\par
{\em{References:}}
\par
[1] U. Leonhardt and P. Piwnicki, Phys. Rev. Lett. {\bf 84}, 822-825 (2000).
\par
[2] U. Leonhardt and P. Piwnicki, Phys. Rev. A {\bf 60}, 4301--4312 (1999).
\par
[3] U. Leonhardt, {\em{Spacetime physics of quantum dielectrics}},
\htmladdnormallink{{\tt{physics/0001064}}}
{http://xxx.lanl.gov/abs/physics/0001064}.
\par
[4] U. Leonhardt,
\htmladdnormallink{
http://www.st-and.ac.uk/\~{}www\_{}pa/group/quantumoptics/media.html}
{http://www.st-and.ac.uk/\~{}www\_{}pa/group/quantumoptics/media.html}
\par
[5] W. Gordon, Ann. Phys. (Leipzig) {\bf 72}, 421 (1923).
\par
[6] W. Unruh, Phys. Rev. Lett. {\bf 46}, 1351 (1981);
Phys. Rev. D {\bf 51}, 2827 (1995).
\par
[7] M. Visser, Class. Quantum Grav. {\bf 15}, 1767 (1998);
see also \htmladdnormallink{{\tt{gr-qc/9311028}}}
{http://xxx.lanl.gov/abs/gr-qc/9311028}.
\par
[8] M. Visser, 
\htmladdnormallink{{\tt{gr-qc/0002011}}}{http://xxx.lanl.gov/abs/gr-qc/0002011}
\par
[9] L. V. Lau, {\em{et al}}, Nature (London) {\bf 397}, 594 (1999).

\vfill
\pagebreak

\section*{\centerline{
``Branification:'' an alternative to compactification}}
\addtocontents{toc}{\protect\smallskip}
\addcontentsline{toc}{subsubsection}{\it
``Branification:'' an alternative to compactification, by Steve Giddings}
\begin{center}
Steve Giddings, University of California at Santa Barbara
\mbox{\ }
\htmladdnormallink{giddings@physics.ucsb.edu}
{mailto:giddings@physics.ucsb.edu}
\end{center}

Recent developments have breathed new life into 
the old idea that the observable Universe is embedded
in a spacetime with extra large or even infinite dimensions. 
This raises the exciting prospect that
Planckian physics could be observed in high-energy accelerators, provides
interesting new techniques to address hierarchy problems in physics, and
could possibly lead to novel phenomena in cosmology and black hole physics.
\par
Obstacles to the viability of such a scenario have 
included explaining why the
matter that we see moves only along the 3+1 dimensional hypersurface, and
explaining the observed gravitational $1/r^2$ force law characteristic of
four dimensions.  Old ideas on
confinement of gauge fields and fermions to a domain wall have been
supplemented with new ones from string theory involving D-branes -- these
address the first issue.  Recall that D-branes
are surfaces which open string ends stick to; if observable matter
consisted of open strings and the Universe was a D3-brane, that could solve
the problem.  But gravity is harder to ``confine'' to a brane-like
structure.
\par
One idea that has been actively pursued by Arkani-Hamed, Dimopoulos, and
Dvali [3] is that the brane is immersed in space with $d$ extra
large but compact dimensions.  If the $d+4$ dimensional fundamental 
Planck mass is $M$,
then the effective four-dimensional Planck mass follows in terms of the
compact volume $V_d$ by an elementary argument from the Einstein-Hilbert action:
$$
{1\over M^{d+2}} \int dV_{d+4} {\cal R} \sim  {V_d \over M^{d+2}} \int
dV_{4} {\cal R} \ ,
$$
giving 
$$M_4^2\sim M^{d+2} V_d\ .\eqno(1)$$  
An alternative explanation of the weakness of gravity is thus not that the
fundamental Planck mass is so big, but rather that the compact volume is
big.  This raises the exciting prospect that the fundamental Planck scale
may be more readily accessible in accelerator experiments, or that the
compact dimensions may be detected through experiments with microgravity
(see the next article in this issue of MOG).
\par
A new variant of this scheme of even more theoretical interest was proposed
by Randall and Sundrum (RS) [4].  In their picture, the brane is
instead the Poincare-invariant boundary of a slice of $4+1$ dimensional
anti-de Sitter space.  RS observed that the negative curvature of anti-de
Sitter space plays a very similar role to that of a compact dimension, and
effectively binds a graviton mode to the brane.  As a result, at low
energies matter living on the brane effectively interacts through
four-dimensional gravity.  The scale at which this ceases to be true, and
the underlying infinite fifth dimension is revealed, is set by the anti-de
Sitter radius, $R$.  The non-compactness of the extra dimension
distinguishes these ``branification'' scenarios from compactification, and
has novel consequences such as the existence of a continuum of
``Kaluza-Klein'' modes.  In analogy to equation (1), we have
$$
M_4^2\sim RM^3\ ,
$$
again raising the possibility that if the anti-de Sitter radius is large
enough, the fundamental Planck scale is commensurately lower and Planckian
or extra-dimensional physics may be much more experimentally accessible.
Variants of the RS proposal have also been considered, involving
either parallel branes in 5 dimensions [5],
which may help with the
hierarchy problem, or intersecting branes in more dimensions.  
\par
Initially there were questions of consistency of this proposal; for example
Chamblin, Hawking, and Reall [6] and others observed the existence of black
holes arising from matter on the brane with infinitely extended horizons
and strong-coupling singularities at the horizon of anti-de Sitter space.
However, they also suggested as a possible resolution that these would
exhibit a Gregory-Laflamme instability resulting in a solution with horizon
confined near the brane.  This expectation was confirmed in the case of a
2+1 dimensional brane by Emparan, Horowitz, and Myers [7], and in a
linearized analysis by Katz, Randall, and the author [8], \
who independently
found that the horizon of such a black hole is shaped like a pancake.
Specifically, its radius along the brane is the familiar $r=2m$, but the
extent transverse to the brane grows only as $R\log m$ with the mass.  
\par
These and other checks in the linearized analysis (properties of
propagators have been worked out in [8]; other linearized analysis appears in 
[1]
support the consistency of RS branification.  Moreover, they raise some
interesting possibilities.  For example, 
we, as four-dimensional observers, would see
processes through their projection onto the brane.  Therefore motion of an
object flying around the pancake-shaped black hole through 
the fifth dimension could
be interpreted by four-dimensional observers as motion into one side of the
horizon and out the other!
\par
More novelties in cosmology arise because of the extra degrees of freedom
associated to motion of the brane or other five-dimensional perturbations
of the metric.  Initially concerns were raised that the Hubble law came out
to be $H\propto \rho$, but more recent work [9,10] 
has shown that in the
presence of extra dynamics that stabilizes the brane's motion we recover
the familiar $H\propto \sqrt\rho$.  More subtle consequences for early
Universe physics are being explored, and there have been suggestions that
these and related scenarios address the cosmological constant
problem [12,13,14]
\par
Finally, the proper setting for branification proposals is presumably string
theory, and direct connection has been made to the celebrated AdS/CFT
correspondence by Maldacena, Witten, Gubser [2]
and [8].  In particular,
H. Verlinde [11] has given a closely related proposal within string theory
compactified (or perhaps noncompactified?) on a noncompact manifold with
an AdS region.  Verlinde's scenario deserves more close scrutiny.
\par
Beyond the need to extend understanding of examples of branification in string
theory, a number of interesting problems remain both in phenomenology (with
a realistic model in hand, what would be the first observable consequence
of this picture?); in cosmology, black hole physics and other aspects of 
the gravitational dynamics in its
subtle interplay between four and five dimensions; and finally, with luck,
in experiment.
\par

{\em{References:}}
\par
[1] J. Garriga and T. Tanaka, ``Gravity in the brane world,'' 
\htmladdnormallink{{\tt{hep-th/9911055}}}
{http://xxx.lanl.gov/abs/hep-th/9911055}. 
\par
[2]S.S. Gubser, ``AdS/CFT and gravity,'' 
\htmladdnormallink
{{\tt{hep-th/9912001}}}
{http://xxx.lanl.gov/abs/hep-th/9912001}.
\par
[3] N. Arkani-Hamed, S. Dimopoulos, and G. Dvali, ``The hierarchy
problem and new dimensions at a millimeter,'' 
\htmladdnormallink
{{\tt{hep-ph/9803315}}}
{http://xxx.lanl.gov/abs/hep-ph/9803315}
Phys. Lett. {\bf B429}  263 (1998); 
``Phenomenology, 
astrophysics and cosmology of theories with submillimeter dimensions and
TeV scale quantum gravity,'' 
\htmladdnormallink
{{\tt{hep-ph/9807344}}}
{http://xxx.lanl.gov/abs/hep-ph/9807344}, {\sl Phys. Rev.} {\bf D59}:086004
(1999).
\par
[4]L. Randall and R. Sundrum, ``An alternative to
compactification,'' 
\htmladdnormallink
{{\tt{hep-th/9906064}}}
{http://www.lanl.gov/abs/hep-th/9906064}, Phys. Rev. Lett. 83 (99) 4690.
\par
[5] J. Lykken and L. Randall, ``The shape of gravity,''
\htmladdnormallink
{{\tt{hep-th/9908076}}}
{http://xxx.lanl.gov/abs/hep-th/9908076}.
\par
[6] A. Chamblin, S.W. Hawking, and H.S. Reall, ``Brane-world black
holes,'' 
\htmladdnormallink
{{\tt{hep-th/9909205}}}
{http://xxx.lanl.gov/abs/hep-th/9909205}.
\par
[7] R. Emparan, G.T. Horowitz, and R.C. Myers, ``Exact description of
black holes on branes,'' 
\htmladdnormallink
{{\tt{hep-th/9911043}}}
{http://xxx.lanl.gov/abs/hep-th/9911043}.
\par
[8] S.B. Giddings, E. Katz, and L. Randall, ``Linearized gravity in
brane backgrounds,'' (to appear); for preliminary accounts see 
S.B. Giddings, talk at ITP Santa Barbara Conference ``New
dimensions in field theory and string theory,''
and 
L. Randall, talk at Caltech/USC conference ``String theory at
the millennium,''\\
\htmladdnormallink{http://www.itp.ucsb.edu/online/susy\_c99/giddings/}
{http://www.itp.ucsb.edu/online/susy\_c99/giddings/} \\
\htmladdnormallink
{http://quark.theory.caltech.edu/people/rahmfeld/Randall/fs1.html}
{http://quark.theory.caltech.edu/people/rahmfeld/Randall/fs1.html}.
\par
[9] C. Csaki, M. Graesser, L. Randall, and J. Terning, ``Cosmology
of brane models with radion stabilization,'' 
\htmladdnormallink
{{\tt{hep-ph/9911406}}}
{http://xxx.lanl.gov/abs/hep-ph/9911406}.
\par
[10] P. Kanti, I.I. Kogan, K.A. Olive, M. Pospelov, 
``Single brane cosmological solutions with a stable compact extra
dimension,'' 
\htmladdnormallink
{{\tt{hep-ph/9912266}}}
{http://xxx.lanl.gov/abs/hep-ph/9912266}.
\par
[11] H. Verlinde, ``Holography and compactification,''
\htmladdnormallink
{{\tt{hep-th/9906182}}}
{http://xxx.lanl.gov/abs/hep-th/9906182}.
\par
[12] J. de Boer, E. Verlinde, H. Verlinde, ``On the holographic
renormalization
group'',\\ 
\htmladdnormallink
{{\tt{hep-th/9912012}}}
{http://xxx.lanl.gov/abs/hep-th/9912012};
E. Verlinde and H. Verlinde, ``RG flow, gravity and the cosmological
constant,'' 
\htmladdnormallink
{{\tt{hep-th/9912018}}}
{http://xxx.lanl.gov/abs/hep-th/9912018};
E. Verlinde, ``On RG flow and the cosmological constant,'' 
\htmladdnormallink
{hep-th/9912058}
{http://xxx.lanl.gov/abs/hep-th/9912058}.
\par
[13] N. Arkani-Hamed, S. Dimopoulos, N. Kaloper, and R. Sundrum, 
``A small cosmological constant from a large extra dimension,''  
\htmladdnormallink
{{\tt{hep-th/0001197}}}
{http://xxx.lanl.gov/abs/hep-th/0001197}.
\par
[14]S. Kachru, M. Schulz, and E. Silverstein, 
``Self-tuning flat domain walls in 5-d gravity and string theory,'' 
\htmladdnormallink
{{\tt{hep-th/0001206}}}
{http://xxx.lanl.gov/abs/hep-th/0001206}.

\vfill
\pagebreak

\section*{\centerline {
Current Searches for non-Newtonian Gravity}\\
\centerline{at Sub-mm Distance Scales}}
\addtocontents{toc}{\protect\smallskip}
\addcontentsline{toc}{subsubsection}{\it
Searches for non-Newtonian Gravity at Sub-mm Distances, by Riley Newman}
\begin{center}
Riley Newman, University of California at Irvine\\
\mbox{\ }
\htmladdnormallink{rdnewman@uci.edu}
{mailto:rdnewman@uci.edu}
\end{center}

{\em{Preface:}} The possibility that ``large" compact dimensions may
have a detectable effect on the gravitational force at small distances
has stimulated many new experiments (see the previous article in this
issue of MOG).  A short sketch of activity in
this field seemed desirable -- of interest to general readers, and of
potential use to current practitioners.  I have found writing this to
be a delicate matter, however.  Several groups have understandably
been hesitant to make public their activity or plans at this stage --
I thank those that consented to go public for this report, and respect
the wish of others not to be publicized at this time.
\par
{\em{The Report:}}  
\par
{\em{John Price}} (John.Price@Colorado.edu) and his postdoc Josh Long
at the University of Colorado are developing an apparatus [1] in which
a vibrating reed source mass is driven at the resonant frequency
(approximately 1 kHz) of a  tungsten plate torsion oscillator
separated from it by a gold-plated thin sapphire shield for
electrostatic shielding.  Capacitive readout of the torsion oscillator
amplitude is used.  The system operates now at room temperature, later
to be at 4K. Mass separations from about 0.1 to 1 mm will be explored,
with a target peak sensitivity about 1\% of gravity at a range of
about 0.3 mm, and  equal to gravity at 0.05 mm.
\par
{\em{The Padua Group}} (eg, Giuseppe.Ruoso@lnl.infn.it) measures the
influence of a stainless steel source mass on the resonance frequency
of a piezo-driven silicon cantilever beam monitored by an optical
fiber interferometer. An earlier version of this experiment [2] put a
limit on a non-Newtonian force at a level of $8 \times 10^7$ of
gravity at 0.2 mm.  Analysis of results from the current version are
underway; Ruoso estimates that the current system is capable of
sensitivity at best about $10^6$ of gravity; this may be improved in
future modifications of the experiment.
\par
{\em{John Lipa}} (John.Lipa@stanford.edu) with S. Wang has built an
apparatus at Stanford which, like the current version of the Padua
experiment, searches for frequency pulling of a mechanical resonator
as a function of field source mass position.  A 6.4 mm diameter
tungsten disk is attached to a torsional oscillator with resonant
frequency 145 Hz and Q 1500, driven capacitively with a phase-locked
loop circuit which tracks its resonant frequency.  The source mass is
a 50 mm diameter tungsten disk, moved at intervals of 20 minutes so
that it is alternately 0.1 mm and 1 mm from the test mass.  The
current sensitivity of the system, operating at room temperature,
corresponds to a force magnitude at 0.1 mm less than $10^5$ of
gravity.  A low temperature version of the system is being considered.
\par
{\em{Aharon Kapitulnik}} (ak@loki.stanford.edu) with Tom Kenny is
operating a system at Stanford with the following design: A test mass
is mounted on a cantilever with very low spring constant ($<10^{-4}
N/m$) within electrostatic shielding, with optical fiber
interferometer readout. The source mass is in the form of five squares
of mass of alternating specific weight (Al and W), caused to swing
periodically laterally about 0.4 mm by a bimorph device.  To generate
the force signal of interest, the bimorph oscillates at a frequency
which is one third of the cantilever's resonance frequency, allowing
excellent inertial decoupling.  The range of mass separations to be
explored is expected to be 0.03 - 0.5 mm.  The ultimate sensitivity of
the present apparatus design, at 4.2 K, is expected to be better than
5\% of gravity at 0.08 mm.  Other designs will be explored in the
future.
\par
{\em{Eric Adelberger}} (Eric@gluon.npl.washington.edu) with Blayne
Heckel is doing an experiment at the University of Washington using a
planar torsion balance that sits above a rotating attractor.  The
apparatus is not yet completed.  Eric's group hopes to be able to
probe with good sensitivity force ranges from 0.05 to 2 mm, and
expects to have some results in about a year if unexpected problems
are not encountered.
\par
{\em{Paul Boynton}} (boynton@u.washington.edu), Michael Moore, and
graduate student Micah Ledbetter at the University of Washington are
considering an experiment in which the signature of non-Newtonian
gravity is a torque on a near-planar torsion pendulum suspended above
a near-planar source mass.  The signal torque is manifested as a
second harmonic distortion of the torsional oscillation of the
pendulum.  The source and pendulum masses are each to be made with
opposing halves at a slightly different elevation, in a configuration
which gives a nearly null signal for purely Newtonian gravity. Source
and test masses are to be separated by a conducting membrane in the
gap between them.  The expected sensitivity to an anomalous force is
at a level of 0.25 of gravity at 0.25 mm and $10^{-2}$ of gravity at 1
mm, limited by machining tolerance.
\par
{\em{Ho Jung Paik}} (h\_paik@umail.umd.edu) at the University of
Maryland has proposed to NSF a mm-scale test for non-Newtonian
gravity.  The proposed system uses two magnetically levitated 2.1 g Nb
test masses 11.6 cm apart, with SQUID readout of their differential
motion. Two nearly planar 1.4 kg source masses would be shaped and
positioned so that when they are moved in opposite directions their
Newtonian effect on the differential motion of the test masses is
null.  The opposite motion of the source masses cancels inertial
reaction forces on the apparatus as a whole, easing vibration
rejection requirements for the experiment.  The source masses
positions will be modulated at about 0.1 Hz.  Test masses will be
shielded from source masses and environment by superconducting
shields.  The design sensitivity of the system is $10^{-4}$ of gravity
at 2 mm and $10^{-2}$ of gravity at 0.1 mm.
\par
{\em{Measurements at very short distances}}.  Experiments designed to
measure the Casimir force can in principle constrain non-Newtonian
gravity, but this is made perilous by uncertainties in accounting for
finite conductivity corrections, surface roughness, dirt, etc.  Two
Casimir force measurements have been made recently:
\par
{\em{Steve Lamoreaux}} (lamore@lanl.gov), used a torsion balance [3] at
the University of Washington to measure the force between a 11.3 cm
spherical lens and a quartz plate, both plated with copper and then
gold, exploring a separation range from 0.6 to 10 microns with results
within about 5\% of the Casimir prediction.  This data has been used
by Price and Long [1] and also by Bordag et al. [4] to constrain
non-Newtonian gravity -- the two analyses appear to disagree somewhat.
The figure in [1] suggests a limit of about $10^{5.7}$ of gravity at
100 microns and $10^{7.3}$ of gravity at 10 microns, while a figure in
[4] implies tighter respective limits of about $10^{3.4}$ and
$10^{6.2}$ of gravity.
\par
{\em{Umar Mohideen}} (umar.mohideen@ucr.edu) with Anushree Roy used an
AFM system at UC Riverside to measure [5] the force between a 0.2 mm
polystyrene sphere and a sapphire plate, both aluminum coated, over a
separation range 100 to 500 nm.  The force was measured with an
average statistical precision over this range equal to about 1\% of
the Casimir force at the smallest surface separation, and was found to
be consistent with the Casimir force using theoretical corrections
calculated to date.  Refinements of this work continue.
\par
{\em{Michael George}} (mgeorge@matsci.uah.edu) and his student Lelon
Sanderson at the University of Alabama, Huntsville, are also
conducting AFM measurements, and exploring with theorist Al Fennelly
the possibility of extracting useful limits on non-Newtonian gravity
from these measurements.
\par
{\em{M. Bordag et al.}} (Michael.Bordag@itp.uni-leipzig.de) have
attempted [6] to constrain sub-micron scale anomalous interactions,
using data from Casimir measurements by others.  However, Lamoreaux
believes that reliable tests for non-Newtonian gravity can only be
made for mass separation greater than 5 or 10 microns, because of
uncertainty in corrections at shorter distances.
\par
{\em{Ephraim Fischbach}} (ephraim@physics.purdue.edu) and his
colleagues are exploring ideas for circumventing some of the perils in
very short distance force measurements, for example by comparing
results obtained using different isotopes of the same materials, which
should have identical electronic properties but differing
gravitational interaction.
\par
{\em{Other experiments.}}  There are undoubtably other sub-mm force
experiments underway or planned.  {\em{Mark Kasevich}}
(mark.kasevich@yale.edu) at Yale, for example, indicated that he
didn't mind being mentioned by name and affiliation, but preferred not
to talk in public about his plans which are still somewhat ill
defined.
\par
I hope that this sketch of current activity in short distance gravity
measurements may be helpful in encouraging communication in an
important field.
\par
{\em{References:}}
\par
[1] J.C. Long, H.W. Chan, J.C. Price, Nuclear Physics B {\bf 539}, 23 (1999). 
\par
[2] C. Carugno, Z. Fontana, R. Onofrio, and C. Rizzo, 
Phys. Rev. D {\bf 55}, 6591, (1997).
\par
[3] S.K. Lamoreaux, Phys. Rev. Lett. {\bf 78}, 5 (1997), 
and Phys. Rev. Lett. {\bf 81}, 5475 (1998).
\par
[4] M. Bordag, B. Geyer, G.L. Klimchitskaya, and V.M. 
Mostepanenko, Phys. Rev. D {\bf 58}, 075003 (1998).
\par
[5] A. Roy, C-Y Lin, and U. Mohideen, Phys. Rev. D {\bf 60}, 111101 (1999).
\par
[6] M. Bordag, B. Geyer, G.L. Klimchitskaya, and V.M. 
Mostepanenko, Phys. Rev. D {\bf 60}, 055004 (1999).

\vfill
\pagebreak

\section*{\centerline {
Quiescent cosmological singularities}}
\addtocontents{toc}{\protect\smallskip}
\addcontentsline{toc}{subsubsection}{\it
Quiescent cosmological singularities by Bernd Schmidt}
\begin{center}
Bernd Schmidt, Albert Einstein Institute, Max Planck Society\\
\mbox{\ }
\htmladdnormallink{schmidt@aei-potsdam.mpg.de}
{mailto:schmidt@aei-potsdam.mpg.de}
\end{center}

In 1964 Penrose and Hawking showed that singularities are a general
feature of classes of solutions of
Einstein's field equations. Their result said nothing about the structure
of those singularities. In the
following years much effort was directed to define and analyze
``singularities" of spacetimes.
However, it turned out that without really using the field equations there
were far too many and absurdly
complicated possibilities such that it seemed hopeless to attempt a useful
classification. In 1970
Belinskii, Khalatnikov and  Lifshitz (BKL) gave a heuristic description of
a class of singularities based
on formal expansions of the metric near a singularity. It remained,
however, unclear whether the use of
the field equation together with the formal expansions could be justified.
\par
A recent theorem by L. Andersson and A. Rendall [1] shows
rigorously that in a particular case,
when the matter is described by a scalar field or a stiff perfect fluid
$(p=\mu)$, the BKL picture is
correct. In this particular case there is no oscillatory behavior near
the singularity, i.e. a quiescent
singularity.
\par
I will describe the theorem  and make some remarks about the way it is
proved. The theorem uses the notion
of ``velocity dominated solution"  which I will define first. Suppose we
have a metric
$$
-dt^2+ g_{ab}(t,x^c)dx^a dx^b\ .
\eqno(1)
$$
If we  drop all spatial derivatives in the evolution equations, we obtain
a system of ordinary
differential equations for $g_{ab}(t)$. In the constraints we just drop
the Ricci scalar. This system
and its solution, $^0g_{ab},^0k_{ab}=\partial_tg_{ab}$, is called ``
velocity dominated". These equations
can be integrated completely and the solutions and their singularities can
be described explicitly. One has
in general
$^0k_a^a=(C-t)^{-1}$. We can chose $C=0$ to have the singularity  occur at
$t=0$. Furthermore all mixed
components  $^0k_a^b$ are proportional to $t^{-1}$. At a fixed spatial
point we can simultaneously
diagonalize $^0k_{ab}$ and $^0g_{ab}$ by a suitable choice of frame. The
diagonal components of the
metric in this frame are then proportional to powers of $t$. The equation
for the matter field can also be
integrated with the result $^0\phi(t,x^a)=A(x^a)\log t +B(x^a)$.
\par
Now we can formulate the main theorem:
\par
{\bf Theorem:} {\it Let $S$ be a three-dimensional analytic
manifold and
($^0g_{ab}(t),^0k_{ab}(t),^0\phi(t)$) a $C^\omega$ solution of the
velocity dominated Einstein--scalar
field equations on $S\times (0,\infty)$ such that $t\  ^0k_a^a=-1$ and
each eigenvalue $\lambda$ of
$ -t\ ^0k_a^b$ is positive. Then there exists an open neighborhood
$U$ of $S\times \{0\}$ in $S\times [0,\infty)$ and a unique $C^\omega$
solution $g_{ab},k_{ab}, \phi $ of the Einstein--scalar field equations on
$U\cap (S\times (0,\infty))$, such that for each compact subset $K\subset
S$
there are real positive numbers $\beta$ and  $\alpha^a{}_b$
with $\beta< \alpha^a{}_b$ for which the following estimates hold
uniformly
on $K$:}
$
\ \ 1.\ \ \ \ ^0g^{ac}g_{cb}=\delta ^a{}_b +o(t^{\alpha^a{}_b})
$

$
\ \ 2.\ \ \ \ k^a{}_b=^0k^a{}_b+o(t^{-1+\alpha^a{}_b})
$

$
\ \ 3.\ \ \ \ \phi=^0\phi+o(t^\beta)
$

$
\ \ 4.\ \ \ \ \partial_t\phi=\partial_t\ ^0\phi+o(t^{-1+\beta})
$

\noindent
{\it together with similar estimates for spatial derivatives of $g_{ab}$
and
$\phi$.}
\bigskip

Each velocity dominated solution is approached by a unique solution of the
full equations. Hence, the
singularity structure of the full solution is the same as that of the
velocity dominated solution. There are
indications that conversely, each solution approaches a velocity dominated
solution.  The behavior of the
curvature tensor near the singularity is
$$
R_{ab}R^{ab}=  K(x^a)\ t^{-4}+ \dots
$$
The solution is really singular at $t=0$ and the BKL picture is justified
in the cases considered.
\bigskip
The proof of the theorem relies on a result by Kichenassamy and Rendall
[2]. It concerns a system of the
form ($u=(u^1\dots u^N),\  x=(x^1,\dots x^n)$)

$$
t\ {\partial u\over\partial t}+A(x)\ u = f(t,x,u,u_x)
$$
Under appropriate conditions, this singular equation has a  unique
solution near $t=0$ which is continuous
in $t$ and tends to zero as $t\to 0$. To use this theorem one  rewrites
the field equations as
equations for the ``difference between the solution and the velocity
dominated solution". Hence regular
equations for a singular solution are replaced by a singular equation for
a regular solution.
\par
New mathematical tools allow fresh investigations of the properties of
singularities of solutions of
Einstein's field equations. Hopefully, this technique can also be used to
treat the more
complicated cases of an  oscillatory behavior of the metric near the
singularity.
\par
{\em{References:}}
\par
[1] Andersson, L. and   Rendall, A. D. Quiescent cosmological
singularities.
\htmladdnormallink{{\tt{gr-qc/0001047}}}
{http://xxx.lanl.gov/abs/gr-qc/0001047}.
\par
[2] Kichenassamy, S. and Rendall, A. D. (1998) Analytic description of
singularities in Gowdy spacetimes. Class.
Quantum Grav. 15, 1339--1355

\vfill
\pagebreak
\section*{\centerline {The debut of LIGO II}}
\addtocontents{toc}{\protect\smallskip}
\addcontentsline{toc}{subsubsection}{\it
The debut of LIGO II, by David Shoemaker}
\begin{center}
David Shoemaker, MIT\\
\mbox{\ }
\htmladdnormallink{dhs@ligo.mit.edu}
{mailto:dhs@ligo.mit.edu}
\end{center}

From its beginning, LIGO was foreseen as capable of supporting a
series of ever-improving detectors over a lifetime of many years. The
LIGO I detector, now being installed, is the exciting first step of
this process. LIGO I uses techniques developed and tested in sensitive
prototypes, begun nearly 30 years ago, and will have a sensitivity
several orders of magnitude greater than any of the gravitational wave
detectors which helped lead to its design. Coincident operation of the
three interferometers at the two LIGO observatories will start in
2002, in coordination with the \htmladdnormallink{VIRGO}{
http://www.pg.infn.it/virgo/} [1] detector and the
\htmladdnormallink{GEO-600}{http://www.geo600.uni-hannover.de} [2]
detector.
\par
Simultaneous with this development of LIGO I, plans
for the more ambitious LIGO II are gaining momentum. The
\htmladdnormallink{LIGO Scientific Collaboration}
{http://www.ligo.caltech.edu/LIGO\_web/lsc/lsc.html}
[3] 
has been refining the vision of what technical advances
would constitute a significant step forward and forming the research
plan which can realize these advances; and the LIGO Laboratory has
studied the practicalities of actually fabricating, installing,
commissioning, and observing with a new detector. An introduction to
this vision of LIGO II is presented in this article. Readers
interested in delving further can investigate the 
\htmladdnormallink
{LSC White Paper} 
{http://www.ligo.caltech.edu/docs/T/T990080-00.pdf} [4] and the 
\htmladdnormallink
{Laboratory Conceptual Plan}
{http://www.ligo.caltech.edu/docs/M/M990288-A1.pdf} [5].
\par
A look at the anticipated sensitivity of LIGO I (seen at right in
Figure 1, top curve) shows three regions; the near-vertical line at
low frequencies, a midrange from $~40$ to $~120$ Hz, and a high frequency
region above 120 Hz. Let's look at how the LIGO II design improves the
performance in each of these regions, starting from the high frequency
end. The design we talk about here is a starting point, rather than a
definition of LIGO II; and we have already been alerted to one missing
component in our model, and anticipate greater thermal noise than
indicated in these curves. With that caveat, here are the broad
outlines of what we hope to achieve with LIGO II.
\begin{figure}[ht]
\centerline{\psfig{file=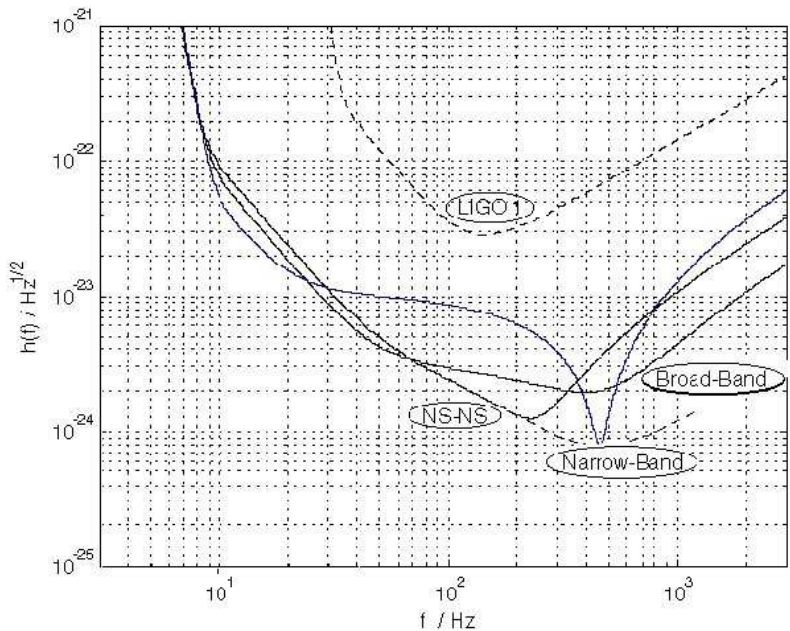,height=10cm}}
\caption{LIGO I and II sensitivities}
\end{figure}
\par
{\em{Shot Noise Dominated Region}}
\par
LIGO I uses 10 W of laser power in a power-recycled Michelson
interferometer with Fabry-Perot arm cavity transducers to sense the
motion of the test masses.  The limit to our ability to sense comes
from the ``shot noise limit"--the (Poisson) statistical fluctuation in
the number of photons arriving at our photodetector makes us uncertain
about the exact position of the test masses. Increasing the laser
power decreases the fractional uncertainty, as the square root of the
laser power, and so an obvious improvement in a second-generation
interferometer is to increase the laser power. The Reference Design
for LIGO II, shown in the
\htmladdnormallink{LSC White Paper} 
{http://www.ligo.caltech.edu/docs/T/T990080-00.pdf}
[4], carries an increase from 10 to 180 W of input
laser power, and also takes advantage of the best optical polishing
and coating to date to allow a lower-loss optical system (and thus a
higher ``recycling gain"). These changes lead to a better sensing of
the test mass motion, and as seen in Figure 1 a much-improved
high-frequency sensitivity. They also require considerable research
and development in optical components: low-noise laser amplifiers,
phase modulators, Faraday isolators, and the means to compensate for
thermal lensing of the interferometer components.
\begin{figure}
\centerline{\psfig{file=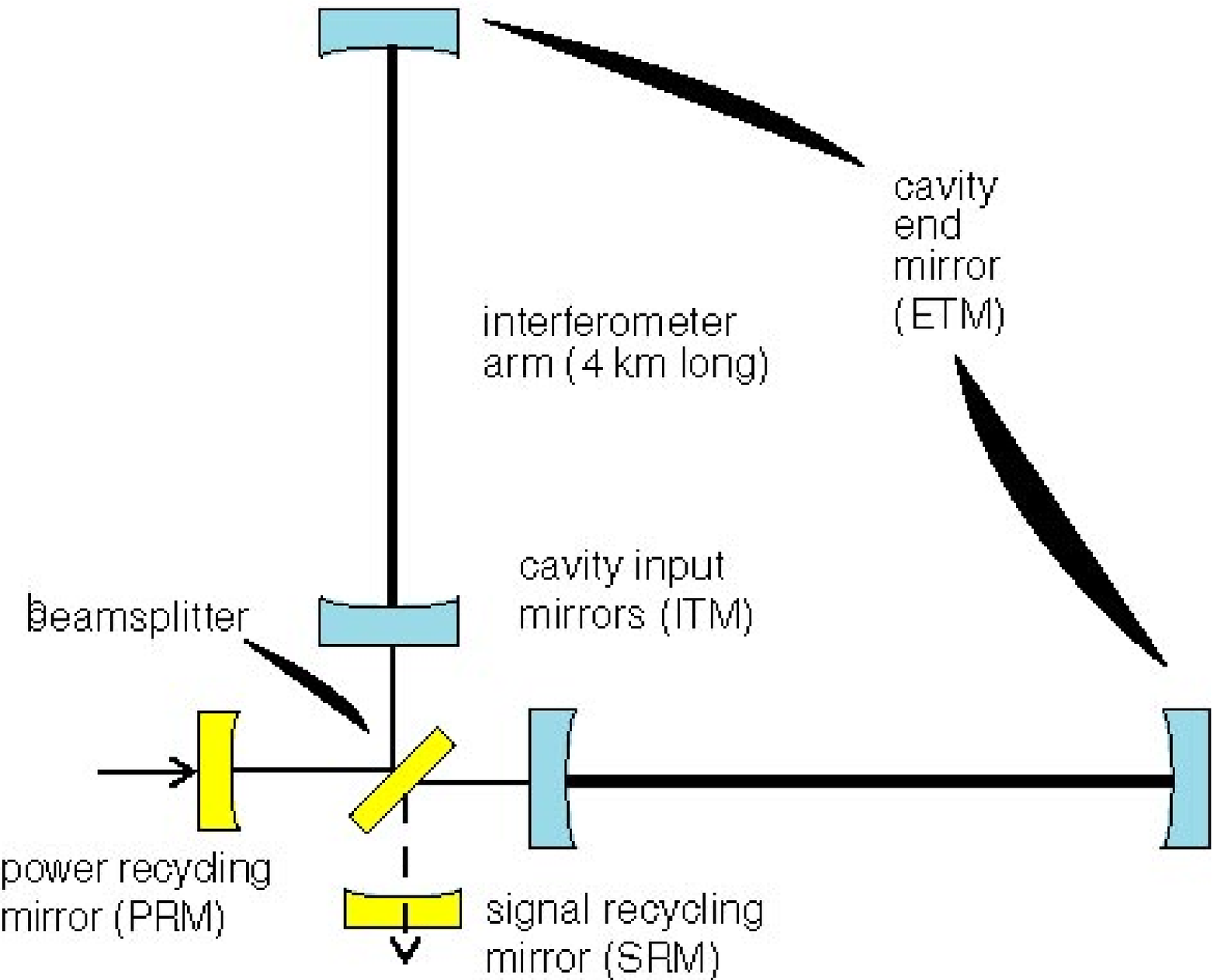,height=10cm}}
\caption{Signal recycling setup}
\end{figure}
However, this is not the only change. There are several curves for
LIGO II shown in Figure 1. This is due to the addition of a Signal
Recycling Mirror, as seen at left in Figure 2. The power recycling
mirror allows unused input power to be ``recycled" into the
interferometer, a technique used in both LIGO I and II.  For LIGO II,
the additional signal recycling mirror can be used either to ``recycle"
(signal recycling), or intentionally ``extract" (resonant sideband
extraction) the actual gravitational-wave induced signal to
selectively increase the sensitivity of the instrument for a specific
signal search. This leads to the collection of curves shown in Figure
1: one can make a frequency response curve which is optimized for,
say, a Neutron Star Binary inspiral event (NS-NS), a broad-band source
such as a Supernova (Broad-Band), or target a specific frequency
(Narrow-Band). These changes are made through sub-wavelength position
changes in the signal recycling mirror position and/or changes in the
effective transmission.
\par
A series of experiments and detailed models have been underway for
some time to both verify the usability of these configurations and to
find a suitable practical form. A significant effort in the Ligo
Science Collaboration Research and Development will be in the
establishment of high-sensitivity prototypes to give confidence in the
design and to test engineering solutions.
\par
We need to delve a little deeper to see a complement to the shot
noise, the radiation pressure noise. Figure 3 at right shows the
aforementioned shot noise contribution to the sensitivity; curve 7
shows the effect of momentum transfer from the photons to the test
masses. The mass motion due to this noise source dominates at low
frequencies, until shot noise takes over at about 100 Hz. This
``buffeting" of the masses {\em{grows}} with the laser power (again as
the square root of the power), and so it becomes clear that an optimum
laser power exists--a power such that the sensing noise at high
frequencies is reduced to an acceptable level, but one where the
low-frequency buffeting of the test masses by radiation pressure is
not so great as to impact the low-frequency performance. We call the
LIGO II design ``a quantum limited interferometer" due to the fact that
at all frequencies the LIGO II sensitivity is limited by the quantum
nature of light.  Since the buffeting is a force, it makes sense that
this noise source falls as $1/f^2$ and that the motion
associated with it becomes smaller if the mass is greater. This leads
us to the second significant change from LIGO I: the test masses are
to be 30 kg, rather than LIGO I's 10 kg, to hold down the radiation
pressure noise (and allow a higher laser power).
\begin{figure}
\centerline{\psfig{file=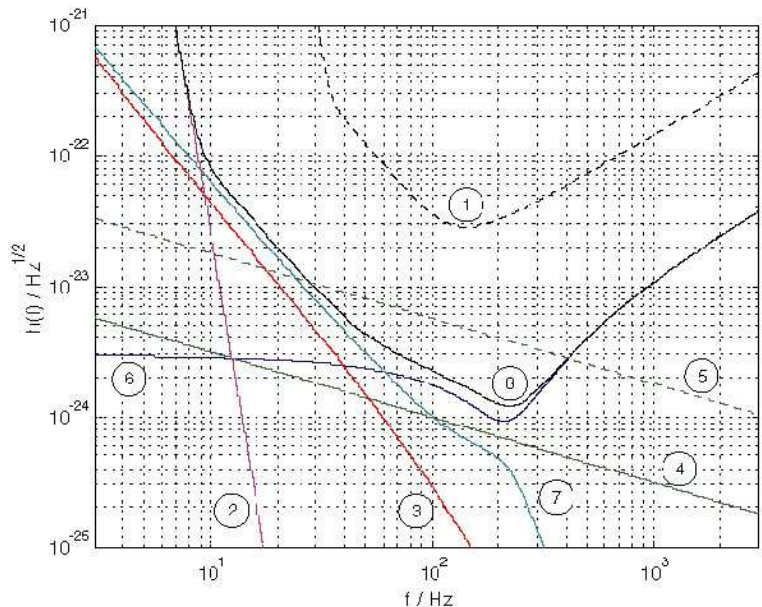,height=10cm}}
\caption{The various components of the noise}
\end{figure}
\par
{\em{Thermal Noise}}
\par
Over a broad range of frequencies, the sensitivity of LIGO I will
be limited by the Brownian motion, and the related noise due to
thermoelastic dissipation, of the test masses. The test masses are in
thermal equilibrium with the surrounding heat bath (at a carefully
regulated 20 degrees Celsius), and thermodynamics tells us that each
mechanical mode of the test masses (and their wire loop pendulum
suspensions, in the case of LIGO I) has kT of energy (where k is
Boltzmann's constant). This energy is expressed as a random motion of
the test mass, where the distribution of the motion as a function of
frequency is determined by details of the losses which limit the
mechanical Q of the system (test mass or suspension). To reduce this
noise, one wants to ``gather" the noise into the peak near the
mechanical resonances (by choosing materials and processes which
maximize the mechanical Q) and place the resonances either below the
frequencies of interest (the pendulum suspension modes, around 1 Hz)
or well above the frequencies of interest (the test mass internal
modes, 10 kHz and higher).
\par
This introduces two very important changes from LIGO I. First, we
are studying the use of sapphire instead of fused quartz for the test
mass material. Sapphire has very low mechanical losses, and also a
high speed of sound and a high density. These are all advantageous for
the thermal noise, and the increased mass is needed for the radiation
pressure noise. However, to obtain sapphire in the size required for a
LIGO test mass (order of 28 cm diameter, 12 cm thickness) and of an
optical quality sufficient for the interferometric sensing, requires a
development effort, but will be rewarded with a much reduced thermal
noise. (Curve 4, Fig 3 shows an estimate for the thermal noise not
including the thermoelastic term, which will increase the level by
factors of 3 to 10 depending upon the frequency). For reference, the
thermal noise for 30 kg fused silica masses is also shown (curve 5);
realizing this alternative test mass material would require physically
large test masses and presents different fabrication challenges.
\par
The second change is to use fused quartz instead of steel wire for the
suspension, and to use a ribbon rather than a simple cylindrical
fiber. Fused quartz is a much lower loss material than wire, and
making a ribbon allows the suspension to be very ``soft" along the
optical path (to store little energy in stiffness of the ribbon
itself, and instead to use gravity as a restoring force for the
pendulum motion) and thus to further reduce the thermal noise from the
fiber (curve 3, Fig 3). The suspension and its design, shown in Figure
4 at right, uses multiple masses and multiple fibers, and is a
contribution from our close collaborators of the German-Scots GEO
group; a similar design will be first tried in the GEO-600
interferometer.
\begin{figure}
\centerline{\psfig{file=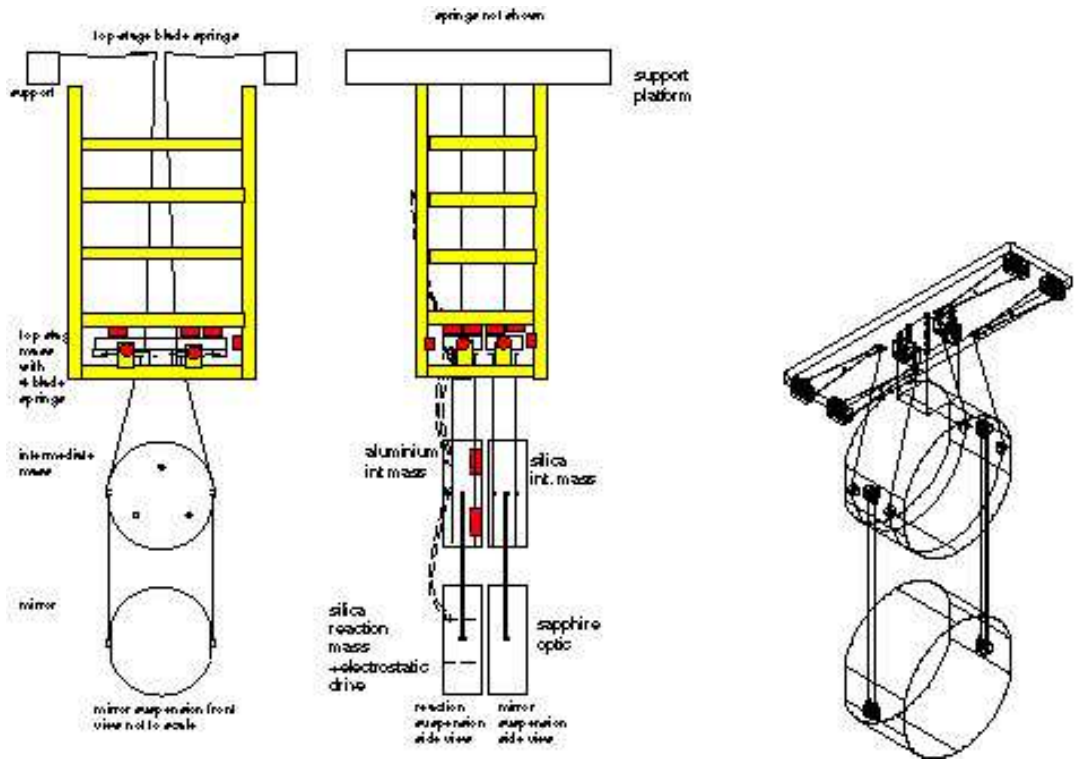,height=10cm}}
\caption{The suspension systems}
\end{figure}
\par
{\em{Seismic Noise}}
\par
The requirement for the attenuation of seismic noise is to make it a
negligible contributor to the overall interferometer performance. Thus
it must be small at all frequencies where other, more difficult and
subtle noise sources (like quantum or thermal noise) are at a level
allowing the observation of probable gravitational wave sources. For
LIGO I, this led to a ``cutoff" or ``brick wall" frequency of 40 Hz--at
all lower frequencies, the thermal noise would have been so great that
no reasonable model for gravitational wave sources would predict
detectable signals.  For LIGO II, due to the much reduced thermal
noise and managed radiation pressure noise, a cutoff frequency of 10
Hz is a good choice (curve 2, Fig 3). This puts the seismic noise
contribution at 10 Hz close to the background due to the Newtonian
background--dynamic changes in the net direction of the gravitational
attraction of the test mass to the earth due to compression and
rarefaction of the nearby earth as normal seismic motion takes
place.
\par
There are two approaches to the seismic attenuation under study. One
uses passive isolation, in a design derived from that used by the
VIRGO project; the other uses servo control techniques to slave the
quiet suspension platform to quiet seismometers.  The final design
must deliver the required reduction in both the seismic noise near 10
Hz as well as fulfilling the very important role of reducing motion
for frequencies 1 Hz and lower as part of the overall control
approach.
\par
To verify the mechanical design of the experiment, a prototype
allowing tests of the suspension and isolation components of LIGO II
is in preparation. The objective, as for the configuration prototypes
mentioned above, is to allow a demonstration of the performance levels
of LIGO II without disturbing observation underway with LIGO I.
\par
{\em{Physics Reach}}
\par
The resulting interferometer (or detector, as interferometers at both
the LIGO Livingston, Louisiana and Hanford, Washington observatories
will be improved) will offer an enormous increase in the sensitivity
to many gravitational wave sources.  In one coarse measure, the strain
sensitivity to broad-band sources in the region of 100 Hz will
increase by a factor of many factors of 10. Because the included
volume of space goes as the cube of the distance, this means we
include many, many times more candidate sources with LIGO II as
compared to LIGO I.  Also, the ``tunability" of the response means
that, as we learn more about specific sources, we can increase our
sensitivity even more dramatically for those sources.  Sources which
might be observable by LIGO I once per year would be observed many
times every day by LIGO II, and the signal-to-noise ratio may allow
detailed studies of the waveform for comparison with numerical models,
leading to better understanding of both astrophysics and of physics in
highly relativistic conditions. The LIGO II detector will be run in
cooperation with the other gravitational-wave detectors to form a
powerful network, permitting the extraction of position, polarization,
and other source parameters from the combined data.
\par
The plan is an ambitious one. We would like to start the replacement
of the LIGO I interferometers with the LIGO II design in 2005 and be
observing before 2007. This schedule will need exquisite preparation
to minimize the ``down time" for observation to a minimum and to assure
that the LIGO II interferometer will perform as designed as quickly as
possible after installation. Close coordination of the Research and
Development leading to a final design is a pre-requisite, and
``all-hands" must be available for the well-rehearsed installation. The
results will be very satisfying of course as a technical
achievement--but more importantly, they will be extraordinarily rich
in astrophysical insights.
\par
{\em{References:}}
\par
[1] \htmladdnormallink{
http://www.pg.infn.it/virgo/}{http://www.pg.infn.it/virgo/} 
\par
[2] \htmladdnormallink{http://www.geo600.uni-hannover.de}
{http://www.geo600.uni-hannover.de}
\par
[3] \htmladdnormallink{http://www.ligo.caltech.edu/LIGO\_web/lsc/lsc.html}
{http://www.ligo.caltech.edu/LIGO\_web/lsc/lsc.html} 
\par
[4] \htmladdnormallink{http://www.ligo.caltech.edu/docs/T/T990080-00.pdf}
{http://www.ligo.caltech.edu/docs/T/T990080-00.pdf}
\par
[5] \htmladdnormallink{http://www.ligo.caltech.edu/docs/M/M990288-A1.pdf}
{http://www.ligo.caltech.edu/docs/M/M990288-A1.pdf}

\vfill
\pagebreak
\vfill
\pagebreak
\section*{\centerline {Is the universe still accelerating?}}
\addtocontents{toc}{\protect\smallskip}
\addcontentsline{toc}{subsubsection}{\it
Is the universe still accelerating?, by Sean Carroll}
\begin{center}
Sean Carroll, University of Chicago\\
\mbox{\ }
\htmladdnormallink{carroll@theory.uchicago.edu}
{mailto:carroll@theory.uchicago.edu}
\htmladdnormallink
{http://pancake.uchicago.edu/{\~{}}carroll/}
{http://pancake.uchicago.edu/\~{}carroll/}
\end{center}

To most cosmologists, it came as something of a surprise when,
in 1998, two groups (the Supernova Cosmology Project
[1]  and the High-Z Supernova Team
[2,3] presented evidence that the expansion
of the universe is accelerating rather than slowing down.
Applied to a Robertson-Walker metric
\begin{equation}
  {\rm d}s^2 = -{\rm d}t^2 + R^2(t)\left[
  {{{\rm d}r^2}\over{1-kr^2}} + r^2 {\rm d}\Omega^2\right]\ ,
  \label{rwmetric}
\end{equation}
Einstein's equations imply the Friedmann equation,
\begin{equation}
  {\dot R}^2 =
  {{8\pi G}\over 3}R^2 \rho - k\ ,
  \label{feq}
\end{equation}
where $R$ is the scale factor, $\rho$ the energy density, and
$k$ the spatial curvature
parameter.  The energy density in a
species of non-relativistic massive particles (``matter'') is given by
the species' rest mass times its number density, and correspondingly
diminishes as $\rho_{\rm M}\propto R^{-3}$ as the number density
becomes increasingly rarified.  In a matter-dominated universe,
then, the right-hand side of (\ref{feq}) is decreasing as
the universe expands, resulting in deceleration.
To provide acceleration ($\ddot R > 0$), the energy density
must decay more slowly than $R^{-2}$; the simplest candidate for
such a source is the cosmological constant $\Lambda$, equivalent
to a ``vacuum'' energy density
\begin{equation}
  \rho_\Lambda = {{\Lambda}\over{8\pi G}}\ ,
\end{equation}
which remains constant as the universe expands.  The supernova
teams have measured the distances to cosmological supernovae by
using the fact that the intrinsic luminosity of Type Ia supernovae,
while not always the same, is closely correlated with their
decline rate from maximum brightness, which can be independently
measured.  Their apparent magnitude then provides an indication of
their distance, and their redshift $z$ (related to the value of the
scale factor $R$ at the time of explosion by $z=R_0/R - 1$) can be
straightforwardly determined from spectroscopic data.  The results
to date favor a positive value of $\rho_\Lambda$.  Along with
constraints on the matter density as derived from dynamical
measurements of galaxies and clusters, and additional constraints
from the anisotropies of the cosmic microwave background, a
consistent picture emerges with $\rho_\Lambda/\rho_{\rm M}
\sim 3$, with the total energy density $\rho_{\rm M}+\rho_\Lambda$
approximately equal to the critical density necessary to solve
(\ref{feq}) with $k=0$.
\par
Despite its excellent fit to the data, such a universe seems
quite unnatural.  For one thing, the implied vacuum energy
$\rho_\Lambda \sim 10^{-10}$~erg/cm$^3$ is less by many orders
of magnitude than any sensible estimate based on particle physics.
For another, $\rho_{\rm M}$ and $\rho_\Lambda$ evolve at different
rates, with $\rho_{\rm M}/\rho_\Lambda \propto R^{-3}$, and it
would seem quite unlikely that they would differ today by a
factor of order unity.  Since any effect which would diminish
the brightness of distant supernovae without noticeably affecting
their spectra could mimic the effects of an accelerating universe,
it is sensible to ask whether these apparently dramatic results can be
explained in terms of conventional astrophysics without invoking new
cosmological phenomena.  The most plausible candidates for such
effects are evolution of the supernova population from high to
low redshifts, and obscuring dust between us and the high-redshift
objects.  Both possibilities are being carefully investigated.
\par
Type Ia supernovae are thought to result from thermonuclear
explosions of white dwarfs which have reached the Chandrasekhar
limit.  Therefore, they can occur in a wide variety of
environments, and a simple argument against evolution is that
the high-redshift environments, while chronologically younger,
should be a subset of all possible low-redshift environments,
which include regions that are ``young'' in terms of chemical
and stellar evolution.  Nevertheless, even a small amount of
evolution could ruin our ability to reliably constrain
cosmological parameters [4].  In their original
papers [1,2,3], the supernova teams
found impressive consistency in the spectral and photometric
properties of Type Ia supernovae over a variety of redshifts
and environments ({\it e.g.}, in elliptical vs.\ spiral
galaxies).  More recently, however, Riess {\it et al.}\
[5]  have presented tentative evidence for a
systematic difference in the properties of high- and low-redshift
supernovae, claiming that the risetimes (from initial explosion
to maximum brightness) were higher in the high-redshift
events.  It is not immediately clear that such a difference
is relevant to the distance determinations; first, because
the risetime is not used in
determining the absolute luminosity at peak brightness, and
second, because a process which only affects the very early
stages of the light curve is most plausibly traced to differences
in the outer layers of the progenitor, which may have a
negligible affect on the total energy output.  Nevertheless,
any indication of evolution brings into question the fundamental
assumptions behind the entire program.  However, Aldering
{\it et al.}\ [6] have argued that the discrepancy
in risetimes goes away once one properly takes into account
correlations in the uncertainties of the light curve fit
parameters.  In that case, all of the data presently available
are consistent with no evolution of any sort between high
and low redshifts.  It is clearly important to improve both
our empirical and theoretical understanding of the high-redshift
supernovae, but to date there is no compelling reason to doubt
the distance determinations (and cosmological conclusions) of
the original studies.
\par
Other than evolution, obscuration by dust is the leading concern
about the reliability of the supernova results.  Ordinary
astrophysical dust does not obscure equally at all wavelengths,
but scatters blue light preferentially, leading to the
well-known phenomenon of ``reddening''.  Spectral measurements
by the two supernova teams reveal a negligible amount of reddening,
implying that any hypothetical dust must be a novel ``grey''
variety.  This possibility has been investigated by a number
of authors [7]
These studies have found that even grey dust is highly constrained
by observations: first, it is likely to be intergalactic rather
than within galaxies, or it would lead to additional dispersion
in the magnitudes of the supernovae; and second, intergalactic dust
would absorb ultraviolet/optical radiation and re-emit it at
far infrared wavelengths, leading to stringent constraints from
observations of the cosmological far-infrared background.
Moreover, even relatively grey dust would inevitably lead to
some reddening, and recent near-infrared observations of a
high-reshift supernova [8] have failed to find any
evidence for such an effect.  Thus,
while the possibility of obscuration has not been entirely
eliminated, it requires a novel kind of dust which is already
highly constrained (and may be convincingly ruled out by
further observations).
\par
Meanwhile, measurements of the anisotropy spectrum of the
cosmic microwave background continue to improve.  Two groups
[9] have reported measurements on the angular
scale of the first ``Doppler peak'', whose location is tied to
the total energy density of the universe.  Both experiments
provide independent evidence that the energy density is approximately
equal to the critical density of a spatially flat universe;
along with increasing confidence that ordinary matter
constitutes approximately $30\%$ if the critical density, this
provides additional support for the existence of a positive cosmological
constant.  Data to come in the near future, from satellite,
ground-based, and balloon-borne experiments, will test this scenario
to much greater precision.  Measurements of additional supernovae at
even higher redshifts have the potential of separating out the
effects of evolution and extinction from those of cosmology; along
with continued ground-based and Space Telescope observations,
a dedicated satellite has been proposed [10] which could
observe 2000 high-redshift supernovae per year.  Our best current
understanding, therefore, continues to favor an accelerating
universe, and in a short while the case could be nailed down
to a near certainty; in that case the task of theorists to
explain a small but nonzero vacuum energy will become especially
urgent.
\par
{\em{References:}}
\par
[1] S. Perlmutter {\it et al.} [Supernova Cosmology
Project Collaboration],
Astrophys. Journ. {\bf 517}, 565
(1999); \htmladdnormallink{{\tt{astro-ph/9812133}}}
{http://xxx.lanl.gov/abs/astro-ph/9812133}.
\par
[2]
B.~P.~Schmidt {\it et al.} [Hi-Z Supernova Team Collaboration],
Astrophys. Journ. {\bf 507}, 46 (1998);
\htmladdnormallink{{\tt{astro-ph/9805200}}}
{http://xxx.lanl.gov/abs/astro-ph/9805200}.
\par
[3]
A.G. Riess {\it et al.} [Hi-Z Supernova Team Collaboration],
Astron. Journ.
{\bf 116}, 1009 (1998); 
\htmladdnormallink{{\tt{astro-ph/9805201}}}
{http://xxx.lanl.gov/abs/astro-ph/9805201}.
\par
[4]
P.~S.~Drell, T.~J.~Loredo and I.~Wasserman,
\htmladdnormallink{{\tt{astro-ph/9905027}}}
{http://xxx.lanl.gov/abs/astro-ph/9905027}.
\par
[5]
A.~G.~Riess, A.~V.~Filippenko, W.~Li and B.~P.~Schmidt,
Astron. Journ. {\bf 118}, 2668 (1999);
\htmladdnormallink{{\tt{astro-ph/9907038}}}
{http://xxx.lanl.gov/abs/astro-ph/9907038}.
\par
[6] 
G. Aldering, R. Knop and P. Nugent,
\htmladdnormallink{{\tt{astro-ph/0001049}}}
{http://xxx.lanl.gov/abs/astro-ph/0001049}.
\par
[7]
A.  Aguirre,
Astrophys. Journ. {\bf 512}, L19 (1999);
\htmladdnormallink{{\tt{astro-ph/9811316}}}
{http://xxx.lanl.gov/abs/astro-ph/9811316};
Astrophys. Journ. {\bf 525}, 583 (1999);
\htmladdnormallink{{\tt{astro-ph/9904319}}}
{http://xxx.lanl.gov/abs/astro-ph/9904319};
A. Aguirre and Z.~Haiman,
\htmladdnormallink{{\tt{astro-ph/9907039}}}
{http://xxx.lanl.gov/abs/astro-ph/9907039};
J.T. Simonsen and S. Hannestad,
Astron. Astrophys. {\bf 351}, 1 (1999);
\htmladdnormallink{{\tt{astro-ph/9909225}}}
{http://xxx.lanl.gov/abs/astro-ph/9909225};
T.~Totani and C.~Kobayashi,
Astrophys. Journ. {\bf 526}, L65 (1999);
\htmladdnormallink{{\tt{astro-ph/9910038}}}
{http://xxx.lanl.gov/abs/astro-ph/9910038}.
\par
[8]
A.G. Riess {\it et al.},
\htmladdnormallink{{\tt{astro-ph/0001384}}}
{http://xxx.lanl.gov/abs/astro-ph/0001384}.
\par
[9]
A.~D.~Miller {\it et al.}, Astrophys. Journ. {\bf 524}, L1 (1999);
\htmladdnormallink{{\tt{astro-ph/9906421}}}
{http://xxx.lanl.gov/abs/astro-ph/9906421};
A. Melchiorri {\it et al.},
\htmladdnormallink{{\tt{astro-ph/9911445}}}
{http://xxx.lanl.gov/abs/astro-ph/9911445}.
\par
[10]
See the web page at 
\htmladdnormallink{http://snap.lbl.gov/}
{http://snap.lbl.gov/}

\vfill
\pagebreak

\section*{\centerline {
Journ\' ees Relativistes Weimar 1999}}
\addtocontents{toc}{\protect\medskip}
\addtocontents{toc}{\bf Conference reports:}
\addtocontents{toc}{\protect\medskip}
\addcontentsline{toc}{subsubsection}{\it  
Journ\' ees Relativistes Weimar 1999, by Volker Perlick}
\begin{center}
Volker Perlick, TU Berlin and 
Albert Einstein Institute, Max Planck Society\\
\htmladdnormallink{vper0433@itp0.physik.tu-berlin.de}
{mailto:vper0433@itp0.physik.tu-berlin.de}
\end{center}

The ``Journ\' ees Relativistes'' started as 
a regular meeting of French relativists several 
decades ago. Over the years, they have grown 
into a series of international conferences held 
in various Western-European cities. The most 
recent meeting in this series took place in 
Weimar, Germany, from September 12 to September 
17, 1999. The main organizers were G.~Neugebauer 
(Chairman) and R.~Collier (Secretary) from the 
relativity group in Jena. The conference was
sponsored by the {\em{Max Planck Society}}, by 
the {\em{Deutsche Forschungsgemeinschaft}}, by 
the {\em{Ministry for Science, Research, and 
Culture\/}} of the state of Thuringia, Germany,
and by the {\em{Friedrich Schiller University\/}} 
at Jena. 
\par
Weimar is a city of approximately 60,000 
inhabitants, situated some 20 kilometers west 
of Jena in the state of Thuringia. The meeting 
took place in a hotel at the outskirts of Weimar 
so that most participants stayed together not 
only during the day but also during the evening. 
As a consequence, there was ample time for vivid 
discussions in a pleasant atmosphere. On Thursday 
afternoon there was no scientific program; instead, 
everyone had the opportunity to visit the city, 
either with an organized tour or on his or her own. 
Weimar is best known for its cultural tradition, 
given the fact that, among others, J.~S.~Bach, 
J.~W.~Goethe, F.~Schiller, and F.~Liszt spent at 
least some years in this city and left their traces 
in various places. For this reason, Weimar is visited 
by a large number of tourists every year. In the 
year of the conference this number was even higher 
than usual because Weimar was nominated ``European 
City of Culture 1999'' by the European Union. 
Incidentally, the same fact had a somewhat unwanted 
impact on the beer prices.
\par
This meeting in Weimar clearly demonstrated the 
international character of the ``Journ\' ees 
Relativistes''. It was attended by participants 
not only from Western Europe but also from 
Eastern-European countries, from various parts of
the former Soviet Union, and from both Americas. 
The total number of participants came up to almost 
100 which was even slightly beyond the seating 
capacity of the lecture room.
\par
Following the tradition of earlier meetings in this 
series, the conference covered all aspects of 
general relativity. The scientific program was 
divided into morning sessions with invited plenary
lectures of 60 minutes or of 30 minutes, two parallel 
afternoon sessions with contributed talks of 30 
minutes, and poster sessions. In the following 
I give a brief overview on the morning sessions.
\par
The conference started with a welcome address by 
G.~Neugebauer (Jena) and a speech by R.~Kerner 
(Paris) honoring the late Andr\' e Lichnerowicz. 
The first scientific talk was by Y.~Choquet-Bruhat 
(Paris) on the so called ``null condition'' and its 
relevance in view of the Christodoulou-Klainerman
result on the global existence of solutions to 
Einstein's vacuum field equation which are close 
to Minkowski space. Classes of solutions to 
Einstein's field equation were investigated also 
in the following talks. H.~Friedrich (Golm) 
considered asymptotically flat solutions and
showed how to calculate some asymptotic quantities 
near spacelike infinity. J.~Bi\v c\' ak (Prague) 
reviewed some recent developments in the investigation 
of radiative spacetimes. D.~Kramer (Jena), stepping 
in as a plenary speaker for Lee Lindblom who could 
not attend the meeting, presented an axially-symmetric 
gravitational wave solution. Z. Perj\' es (Budapest) 
talked about general properties of rotating perfect 
fluid solutions and on strategies of finding such 
solutions that may serve as models of rotating stars. 
For the idealized case of a rigidly rotating disk of 
dust, this problem was solved by Neugebauer and Meinel 
a few years ago. G.~Neugebauer (Jena) in his talk 
elucidated that this was possible by viewing the
whole problem as a boundary value problem for the 
exterior vacuum region and rewriting this as a 
Riemann-Hilbert problem. R.~Meinel (Jena) in his 
talk discussed the properties of this disk-of-dust 
solution in a common setting with the Kerr solution. 
(There was also a video presentation in one of the 
afternoon sessions by M.~Ansorg (Jena) and 
D.~Weiskopf (T\" ubingen) visualizing the optical 
appearance of the Neugebauer-Meinel disk to an
outside observer.)
\par
Various aspects of black holes were at the center of 
a second group of talks. Th.Damour (Paris) spoke
on a correspondence between self-gravitating string states 
and Schwarzschild black holes. D.~Brill (Maryland) 
presented solutions to the $(2+1)$-dimensional source-free 
Einstein equation that are constructed by gluing
together several black-hole configurations. W.~Israel 
(Victoria) discussed some aspects of the thermodynamics 
of spinning black holes. G. Sch{\"a}fer (Jena) reported 
on his results with P.~Jaranowski, presenting the 
dynamics of binary black-hole systems to within
3rd post-Newtonian approximation. B.~Br\" ugmann 
(Golm) gave a status report on results of the 
so-called Grand Challenge Alliance, aiming at 
numerically investigating dynamical processes
such as binary black-hole mergers. 
\par
Quantum field theory on a classical spacetime background 
was the topic of the talk by V.~Belinski (Moscow) who 
critically discussed various derivations of the Unruh 
effect. In addition, there were two talks aiming at 
quantizing the gravitational field itself. S.~Deser 
(Brandeis) talked about ultraviolet divergences 
in quantum (super-)gravity, indicating the necessity 
of stringlike, nonlocal, extensions. A.~Ashtekar (Penn 
State) considered spacetimes with ``isolated horizons'', 
a notion which generalizes the event horizons known from 
the theory of static black holes, and the non-perturbative 
quantization of such objects. 
\par
Another group of talks can be summarized under the 
heading ``cosmology''. This includes a review on the 
microwave background radiation and its anisotropies by 
N.~Deruelle (Paris) and a talk on how to find limits for 
the cosmological parameters with the help of gravitational 
lens statistics by N.~Straumann (Z\"urich). The foundations
of gravitational lens theory from a spacetime perspective, 
concentrating on the geometry of light cones, were 
discussed by J.~Ehlers (Golm). There were two more talks
with a relation to cosmology. B.~Carter (Paris) discussed 
various aspects of cosmic strings, and M.~Demia\' nski 
(Warsaw) reviewed the history of the gravitational constant.
\par
The program was rounded out by one plenary talk on 
experimental aspects of gravity. H.-P.~Nollert (T\" ubingen)
gave an overview on the prospects of gravitational wave 
astronomy.
\par
Further information can be found on the conference homepage
\begin{center}
\htmladdnormallink{
{\tt{http://www.tpi.uni-jena.de/tpi/journees-relativistes.html}}}
{http://www.tpi.uni-jena.de/tpi/journees-relativistes.html}
\end{center}
Written versions of all presentations (invited talks, 
contributed talks, and posters) that survive a refereeing 
process will be published in a double issue of {\em{
Annalen der Physik $($Leipzig}}).

\vfill
\pagebreak

\section*{\centerline {
The 9th Midwest Relativity Meeting}}
\addtocontents{toc}{\protect\medskip}
\addcontentsline{toc}{subsubsection}{\it  
The 9th Midwest Relativity Meeting, by Thomas Baumgarte}
\begin{center}
Thomas Baumgarte, University of Illinois at Urbana-Champaign\\
\htmladdnormallink{thomas@astro.physics.uiuc.edu}
{mailto:thomas@astro.physics.uiuc.edu}
\end{center}

The 9th Midwest Relativity Meeting was hosted by Stu Shapiro, Thomas
Baumgarte and the Illinois Relativity Group at the Department of Physics
of the University of Illinois at Urbana-Champaign on November 12 \&
13, 1999.  With about 80 participants and over 50 presentations it was
the largest Midwest Relativity Meeting so far.  A list of
participants, program, and transparencies of all the talks can be found
at the conference's website \htmladdnormallink{{\tt{http://www.pws.uiuc.edu/groups/relativity/MRM9/}}}
{http://www.pws.uiuc.edu/groups/relativity/MRM9/}.
\par
In the tradition of the regional meetings in the US there were no
parallel sessions, and all talks were limited to 10 minutes, plus 5
minutes for questions.  The talks were grouped into nine sessions,
covering gravitational waves, numerical relativity (two sessions),
energy and entropy, relativistic astrophysics, perturbative methods,
cosmology, mathematical relativity and quantum gravity, and
mathematical and theoretical issues.  In the following I will briefly
mention some of the most interesting contributions, and I apologize to
all those speakers who I have left out.
\par
John Friedman started off the meeting with a summary of recent work on
unstable r-modes in rotating neutron stars.  Fred Lamb later picked up
the story and reported on possible limits to r-mode instabilities due
to magnetic fields.  After a brief update on the current status of the
LISA project, Peter Bender discussed the prospect of detecting
gravitational waves from massive black holes and their coalescence.
Bill Hiscock discussed MACHOs in the Galactic halo as sources of low
frequency gravitational waves, which may also be detectable by LISA.
\par
A number of interesting new results were presented in the two sessions
on numerical relativity, demonstrating that numerical relativity is
now able to address longstanding, three-dimensional problems in
gravitational physics and astrophysics.  Stu Shapiro presented results
on the stability and collapse of relativistic, rotating neutron stars.
Masaru Shibata discussed fully relativistic simulations of binary
neutron star mergers.  His results suggest that the merger may lead to
a very massive neutron star as opposed to a prompt collapse to a black
hole.  Thomas Baumgarte showed how such ``hyper-massive'' neutron star
can be stabilized against collapse by virtue of differential rotation.
Walter Landry demonstrated that a particular implementation of a
higher order diffusion term can stabilize the otherwise unstable
numerical evolution of the ADM equations.  Simonetta Fritelli showed
how recent, conformally decomposed versions of the ADM equations,
which have shown much better numerical behavior than the original ADM
equations, can be cast into a first-order well-posed form.  Wai-Mo
Suen, Mark Miller and other members of the Washington University group
presented updates on the status of the NASA Neutron Star Grand
Challenge Project, including simulations of coalescing neutron stars.
Roberto Gomez discussed horizon data for black hole collisions, and
Mijan Huq presented simulations of grazing collisions of black holes.
\par
Bob Wald presented a generalization of the ``Bousso bound'' (or
``holographic bound'') on the entropy flux through a null
hypersurface.  Robert Mann showed how the entropy, energy and angular
momentum of Misner strings emerge from boundary terms of the
gravitational action in the AdS/CFT correspondence.  Matt Visser
demonstrated how certain quantum effects and even some classical
systems can lead to violations of all the energy conditions of general
relativity, and Carlos Barcelo discussed some of the consequences.
\par
Shmulik Balberg discussed the effect of accretion onto black holes in
core-collapse supernovae on the supernova light curve.  In particular,
he pointed out that for SN1997D in NGC1536 these effects may well be
observable in the next year.  Draza Markovic presented results on
gravitomagnetic warping modes of inner accretion disks, which may
explain the quasi-periodic X-ray brightness oscillations observed in
X-ray binaries.
\par
Eric Poisson and Bill Laarakkers discussed how the presence of a
cosmological horizon in Schwarzschild-deSitter spacetimes affects the
radiative falloff of a massless scalar field.
\par
Leonard Parker explained how non-perturbative terms in the vacuum
energy-momen\-tum tensor of a quantized field can cause an
acceleration of the recent expansion of the universe, and Alpan Raval
showed how this model fits current cosmological observations,
including data from high-red\-shift Type Ia supernovae.
\par
I think that it was a very interesting and lively meeting, enjoyable
even for the organizers.  They would especially like to thank the
Department of Physics at the University of Illinois once again for its
generous support of this meeting.  The 10th Midwest Relativity Meeting
will be hosted by Beverly Berger and David Garfinkle at Oakland
University, tentatively scheduled for Oct. 27 \& 28 2000.

\end{document}